\documentclass[a4paper,11pt]{JHEP3}

\usepackage{amssymb}
\usepackage{citesort}

\def\bra#1{\left\langle{#1}\right|}
\def\ket#1{\left|#1\right\rangle}

\author{
Leszek Hadasz\footnote{Alexander von Humboldt Fellow;
    e-mail: hadasz@th.if.uj.edu.pl} \\
    Physikalisches Institut,
    Rheinische Friedrich-Wilhelms-Universit\"at,
    Nu\ss{}allee 12, 53115~Bonn, Germany
    \\
    and \\
    M. Smoluchowski Institute of Physics,
    Jagiellonian University,
    Reymonta 4,
    30-059~Krak\'ow, Poland}
\author{
Zbigniew Jask\'olski\footnote{jask@ift.uni.wroc.pl}\\
    Institute of Theoretical Physics,
    University of Wroc{\l}aw,
    pl. M. Borna, 950-204~Wroc{\l}aw, Poland
    }
\author{
Paulina Suchanek\footnote{suchanek@th.if.uj.edu.pl} \\
    M. Smoluchowski Institute of Physics,
    Jagiellonian University,
    Reymonta 4, 30-059~Krak\'ow, Poland}

\abstract{Four-point super-conformal blocks for the N = 1
Neveu-Schwarz algebra are defined in terms of power series
of the even super-projective invariant.
Coefficients of these expansions are represented
both as sums over poles in
the ``intermediate'' conformal weight and as sums over poles in
the central charge of the algebra.
The residua of these poles are
calculated in both cases.
Closed recurrence relations for the block coefficients
are derived.}

\title{Recursion representation of the Neveu-Schwarz superconformal block}

\preprint{
hep-th/0611266\\
BONN-TH-2006-07\\
IFT UWr 0211/006 }

\begin{document}

\section{Introduction}
Since the appearance of the seminal BPZ work \cite{Belavin:1984vu},
the conformal block has been recognized as one of the basic objects in conformal field theory.  In spite
of the progress achieved over the years its explicit calculation is still one of the most difficult problems in CFT.
Not only the form of a general conformal block is unknown, but also its analytic
properties still have a status of conjectures rather than theorems.
On the other hand, from the practical point of view
there exist very efficient recursive methods of an approximate (with arbitrary precision), analytic determination
of a general 4-point conformal block
\cite{Zamolodchikov:ie,Zamolodchikov:2,Zamolodchikov:3}.
They were used for instance in checking the conformal bootstrap in the Liouville theory with the DOZZ
coupling constants \cite{Zamolodchikov:1995aa}, in study of the $c\to 1$ limit of minimal models
\cite{Runkel:2001ng} or in obtaining new results in the classical geometry of hyperbolic surfaces
\cite{Hadasz:2005gk}. In a more general context of an arbitrary CFT model these methods
allow for efficient  numerical calculations of any 4-point function once the structure constants are known.

Similar problems can be also addressed in the supersymmetric conformal field theories (SCFT).
It seems therefore desirable to develop analogous recursive methods also in this more complicated  case.
The present paper is aimed as a step in this direction. After defining conformal blocks
for the Neveu-Schwarz algebra of the $N=1$ SCFT we give their representations in
the form of a sum over poles in the central charge and
in the ``intermediate'' conformal weight and derive a recurrence relation for
residua of these poles. In the case of the sum over the poles in the central charge one
can easily calculate the leading ($c \to \infty$) term. This yields efficient recursion relations
for the coefficients of the so called $x$-expansion of NS superconformal blocks.

There are two problems which are natural continuations of the present work.
The first one is to extend our constructions and results to the Ramond sector.
The second is to develop recursion relations for the coefficients of the so called $q$-expansion
\cite{Zamolodchikov:2,Zamolodchikov:3}.
The latter problem requires an explicit calculations of
 the asymptotic behavior of  the superconformal block
for large intermediate weight.
We hope to present the solutions to these problems in the near future.

\section{Four-point correlation functions of the Neveu-Schwarz sector }
The superconformal symmetry is generated by a pair of holomorphic currents
$T(z),\, S(z)$ and their anti-holomorphic counterparts ${\overline T}(\bar z),\,{\overline S}(\bar z),$
where $T$ and ${\overline T}$ are components of the energy-momentum tensor while $S$  and ${\overline S}$
have dimensions $(3/2,0)$ and $(0,3/2),$ respectively. The algebra of the modes of $T(z)$ and $S(z)$ is
determined by the OPE-s
\begin{eqnarray}
\label{OPE:TS}
\nonumber
T(z)T(0) & = & \frac{c}{2z^4} + \frac{2}{z^2}T(z) + \frac{1}{z}\partial T(0) + \ldots,
\\
T(z)S(0) & = & \frac{3}{2z^2}S(0) + \frac{1}{z}\partial S(0)+ \ldots,
\\
\nonumber
S(z)S(0) & = & \frac{2c}{3z^3} + \frac{2}{z}T(0) + \ldots\,.
\end{eqnarray}
The space of fields  of superconformal field theory (hereafter SCFT) decomposes onto
the space of fields $\phi_{\rm NS}$ local with respect to $S(z)$, and the
space of the Ramond fields $\phi_{\rm R}$ with the property that the correlation functions
\[
\left\langle
S(z)\phi_{\rm R}(z_1,\bar z_1)\ldots
\right\rangle
\]
(with the dots denoting any other operator insertions) change the sing upon analytic continuation
in $z$ around the point $z=z_1.$ It the present paper we shall discuss only the NS fields.

The locality properties of the NS field allow to write its OPE with $S(z)$ in the form
\[
%\begin{equation}
S(z)\phi_{\rm NS}(0,0) \; = \; \sum\limits_{k\in {\mathbb Z} + \frac12} z^{k-\frac32}S_{-k}\phi_{\rm NS}(0,0).
%\end{equation}
\]
Together with the usual Virasoro generators $L_n$ defined by the OPE
\[
%\begin{equation}
T(z)\phi_{\rm NS}(0,0) \; = \; \sum\limits_{k\in {\mathbb Z}} z^{n-2}L_{-n}\phi_{\rm NS}(0,0),
%\end{equation}
\]
$S_k$ form the Neveu-Schwarz algebra determined by (\ref{OPE:TS}),
\begin{eqnarray}
\label{NS}
\nonumber
\left[L_m,L_n\right] & = & (m-n)L_{m+n} +\frac{c}{12}m\left(m^2-1\right)\delta_{m+n},
\\
\left[L_m,S_k\right] & = &\frac{m-2k}{2}S_{m+k},
\\
\nonumber
\left\{S_k,S_l\right\} & = & 2L_{k+l} + \frac{c}{3}\left(k^2 -\frac14\right)\delta_{k+l}.
\end{eqnarray}

In the space of all NS fields there exist ``super-primary'' fields
$\varphi_{\Delta,\bar\Delta}(z,\bar z)$
with the conformal weights $\Delta, \bar \Delta,$ which satisfy
\begin{eqnarray}
\label{primary}
\nonumber
\left[L_n,\varphi_{\Delta,\bar\Delta}(0,0)\right]
& = &
\left[S_k,\varphi_{\Delta,\bar\Delta}(0,0)\right]
\; = \; 0,
\hskip 1cm
n,k > 0,
\\[6pt]
\left[L_0,\varphi_{\Delta,\bar\Delta}(0,0)\right]
& = &
\Delta\varphi_{\Delta,\bar\Delta}(0,0),
\end{eqnarray}
and similarly for the ``right'' generators $\bar L_n$ and $\bar S_k.$ Each super-primary
field is the ``lowest'' component of the superfield
\begin{equation}
\label{NS:superfield}
\Phi_{\Delta,\bar\Delta}(z,\theta;\bar z,\bar\theta)
\; = \;
\varphi_{\Delta,\bar\Delta}(z,\bar z)
+
\theta\,\psi_{\Delta,\bar\Delta}(z,\bar z)
+
\bar\theta\,\bar\psi_{\Delta,\bar\Delta}(z,\bar z)
-
\theta\bar\theta\,\widetilde\varphi_{\Delta,\bar\Delta}(z,\bar z),
\end{equation}
where
\[
\psi_{\Delta,\bar\Delta}
\; = \;
\left[S_{-1/2},\varphi_{\Delta,\bar\Delta}\right],
\hskip 5mm
\bar\psi_{\Delta,\bar\Delta}
\; = \;
\left[\bar S_{-1/2},\varphi_{\Delta,\bar\Delta}\right],
\hskip 5mm
\widetilde\varphi_{\Delta,\bar\Delta}
\; = \;
\left\{S_{-1/2},\left[\bar S_{-1/2},\varphi_{\Delta,\bar\Delta}\right]\right\},
\]
and $\theta,\bar\theta$ are Grassman (odd) numbers. The superfield (\ref{NS:superfield}) satisfies
the equations:
\begin{eqnarray}
\label{super:comutators}
\nonumber
\left[L_n,\Phi_{\Delta,\bar\Delta}(z,\theta;\bar z,\bar\theta)\right]
& = &
z^n\left[
z\partial_z +(n+1) \left(\Delta+\frac12\theta\partial_\theta\right)
\right]
\Phi_{\Delta,\bar\Delta}(z,\theta;\bar z,\bar\theta),
\\[8pt]
\nonumber
\left[\bar L_n,\Phi_{\Delta,\bar\Delta}(z,\theta;\bar z,\bar\theta)\right]
& = &
\bar z^n\left[
\bar z\partial_{\bar z} +(n+1)\left(\bar\Delta+\frac12\bar\theta\partial_{\bar\theta}\right)
\right]
\Phi_{\Delta,\bar\Delta}(z,\theta,\bar z;\bar\theta),
\\[8pt]
\left[S_k,\Phi_{\Delta,\bar\Delta}(z,\theta;\bar z,\bar\theta)\right]
& = &
z^{k-\frac12}\left[
z\partial_\theta -\theta z\partial_z -\theta(2k+1)\Delta
\right]
\Phi_{\Delta,\bar\Delta}(z,\theta;\bar z,\bar\theta),
\\[12pt]
\nonumber
\left[\bar S_k,\Phi_{\Delta,\bar\Delta}(z,\theta;\bar z,\bar\theta)\right]
& = &
\bar z^{k-\frac12}\left[
\bar z\partial_{\bar\theta} -\bar\theta\bar z\partial_{\bar z} -\bar\theta(2k+1)\bar \Delta
\right]
\Phi_{\Delta,\bar\Delta}(z,\theta;\bar z,\bar\theta).
\end{eqnarray}
Operators $L_0,S_{\pm\frac12},L_{\pm 1}$ (as well as their right counterparts) form a closed
subalgebra of the NS algebra and exponentiate to the group of the ``super-projective'' transformations.
These transformations allow to express three-point function of the primary superfields in the form
\begin{eqnarray}
\nonumber
&&
\hspace*{-1.8cm}
\Big\langle
\Phi_3(z_3,\theta_3;\bar z_3,\bar\theta_3)
\Phi_2(z_2,\theta_2;\bar z_2,\bar\theta_2)
\Phi_1(z_1,\theta_1;\bar z_1,\bar\theta_1)
\Big\rangle
\\
&& = \;
Z_{32}^{\gamma_1}\bar Z_{32}^{\bar \gamma_1}\,
Z_{31}^{\gamma_2}\bar Z_{31}^{\bar \gamma_2}\,
Z_{21}^{\gamma_3}\bar Z_{21}^{\bar \gamma_3}\
\Big\langle
\Phi_1(\infty,0;\infty,0)
\Phi_2(1,\Theta;1,\bar\Theta)
\Phi_3(0,0;0,0)
\Big\rangle,
\end{eqnarray}
where $\gamma_1 = \Delta_1 -\Delta_2 -\Delta_3,\ $  $Z_{12} = z_1-z_2 - \theta_1\theta_2 \equiv z_{12} - \theta_1\theta_2$ etc.,
\[
\Theta \; = \; \frac{1}{\sqrt{z_{12}z_{13}z_{23}}}
\left(\theta_1 z_{23} +\theta_2 z_{31} + \theta_3 z_{12} - \frac12\theta_1\theta_2\theta_2\right),
\]
and
\[
\Phi_3(\infty,0;\infty,0) \equiv \;  \lim_{R\to\infty} R^{2\Delta_3+2\bar\Delta_3}\Phi_3(R,0;R,0).
\]
Thus, in contrast to the non-supersymmetric case, the three point function is determined
by the superconformal symmetry up to
{\em two} independent constants,
\begin{eqnarray}
\label{constant1}
C_{321} & = & \Big\langle\varphi_3(\infty,\infty)\varphi_2(1,1)\varphi_1(0,0)\Big\rangle,
\\[6pt]
\label{constant2}
\widetilde C_{321} & = & \Big\langle\varphi_3(\infty,\infty)\widetilde\varphi_2(1,1)\varphi_1(0,0)\Big\rangle.
\end{eqnarray}
The super-projective invariance also allows to express a general four-point function
through the four-point function of the form
\[
\bra{\Delta_4,\bar\Delta_4} \Phi_3(1,\theta_3; 1,\bar{\theta}_3)
\Phi_2(z,\theta_2; \bar{z},\bar{\theta}_2) \ket{\Delta_1,\bar\Delta_1}.
\]
Expanding the superfields in the fermionic arguments we get with the help of (\ref{NS:superfield}):
\begin{eqnarray*}
&&\hspace*{-20pt}
\bra{\Delta_4,\bar\Delta_4} \Phi_3(1,\theta_3; 1,\bar{\theta}_3)
\Phi_2(z,\theta_2; \bar{z},\bar{\theta}_2) \ket{\Delta_1,\bar\Delta_1} =
\\[4pt]
&&
=\,
\bra{\Delta_4,\bar\Delta_4} \phi_3(1,1) \phi_2(z, \bar{z}) \ket{\Delta_1,\bar\Delta_1}
+
\theta_3 \bar{\theta}_3  \theta_2 \bar{\theta}_2\bra{\Delta_4,\bar\Delta_4} \tilde{\phi}_3(1,1) \tilde{\phi}_2(z, \bar{z}) \ket{\Delta_1,\bar\Delta_1}
\\[4pt]
&&
-\;
\theta_2 \bar{\theta}_2 \bra{\Delta_4,\bar\Delta_4} \phi_3(1,1)\tilde{\phi}_2(z, \bar{z}) \ket{\Delta_1,\bar\Delta_1}
-
\theta_3 \bar{\theta}_3 \bra{\Delta_4,\bar\Delta_4} \tilde{\phi}_3(1,1) \phi_2(z, \bar{z})  \ket{\Delta_1,\bar\Delta_1}
\\[4pt]
&&
+\;
\theta_3  \theta_2  \bra{\Delta_4,\bar\Delta_4} \psi_3(1,1) \psi_2(z,\bar{z}) \ket{\Delta_1,\bar\Delta_1}
+
\bar{\theta}_3 \bar{\theta}_2  \bra{\Delta_4,\bar\Delta_4} \bar{\psi}_3(1,1) \bar{\psi}_2(z, \bar{z})\ket{\Delta_1,\bar\Delta_1}
\\[4pt]
&&
+ \;
\theta_3 \bar{\theta}_2 \bra{\Delta_4,\bar\Delta_4} \psi_3(1,1)\bar{\psi}_2(z, \bar{z}) \ket{\Delta_1,\bar\Delta_1}
+
\bar{\theta}_3  \theta_2 \bra{\Delta_4,\bar\Delta_4} \bar{\psi}_3(1,1) \psi_2(z, \bar{z}) \ket{\Delta_1,\bar\Delta_1}.
\end{eqnarray*}

\section{NS supermodule}
Let $\nu_{\Delta}$ be the highest weight state with respect to the NS superconformal
algebra (\ref{NS})
\begin{equation}
\label{highest}
L_0\nu_{\Delta}=\Delta\nu_{\Delta}\,,
\hskip 5mm
L_m\nu_{\Delta}= S_k\nu_{\Delta} =0,
\hskip 5mm
m\in \mathbb{N},\;k\in \mathbb{N}-\textstyle{1\over 2},
\end{equation}
where $ \mathbb{N}$ is the set of positive integers.
We denote by ${\cal V}^{f}_\Delta$ the free vector space  generated by all vectors
of the form
\begin{equation}
\label{basis}
\nu_{\Delta,KM}
\; = \;
S_{-K} L_{-M}\nu_{\Delta}
\; \equiv \;
S_{-k_i}\ldots S_{-k_1}L_{-m_j}\ldots L_{-m_1}\nu_{\Delta}\,,
\end{equation}
where
 $K = \{k_1,k_2,\ldots,k_i\}\subset \mathbb{N}-{1\over 2} $ and
 $M = \{m_1,m_2,\ldots,m_j\}\subset \mathbb{N}$ are
 arbitrary ordered
 sets of  indices
\[
k_i < \ldots < k_2 < k_1 ,
\hskip 1cm
m_j \leq \ldots \leq m_2 \leq m_1,
\]
such that $
|K|+|M|= k_1+\dots+k_i+m_1+\dots+m_j = {f}
$.

The ${1\over 2}\mathbb{Z}$-graded representation of the NS superconformal algebra
determined on the space
$$
{\cal V}_\Delta\
=
\hspace*{-3mm}
\bigoplus\limits_{{f}\in {1\over 2}\mathbb{N}\cup\{0\}}
\hspace*{-1mm}
{\cal V}^{f}_\Delta\,,
\hskip 5mm
{\cal V}^0_\Delta=\mathbb{C}\, \nu_\Delta\,,
$$
by the relations
(\ref{NS}) and (\ref{highest}) is called the NS supermodule
of the highest weight $\Delta$ and the central charge $c$
(to avoid making the notation overloaded we omit the subscript $c$ at $\cal V$).
Each ${\cal V}^{f}_\Delta$ is an eigenspace of $L_0$ with the eigenvalue
$\Delta +{f}$. The space ${\cal V}_\Delta$ has also a natural $
\mathbb{Z}_2$-grading:
$$
{\cal V}_\Delta
\; = \;
{\cal V}^+_\Delta \oplus
{\cal V}^-_\Delta\,,
\hskip 5mm
{\cal V}^+_\Delta
\; = \hskip -5pt
\bigoplus\limits_{m\in \mathbb{N}\cup \{0\}}
{\cal V}^m_\Delta\,,
\hskip 5mm
{\cal V}^-_\Delta
\; = \hskip -5pt
\bigoplus\limits_{k\in  \mathbb{N}-{1\over 2}}
{\cal V}^k_\Delta\,,
$$
where ${\cal V}^\pm_\Delta$ are eigenspaces of the
parity operator $(-1)^F= (-1)^{2(L_0-\Delta)}$.

The tensor product ${\cal V}_\Delta \otimes \bar{\cal V}_{\bar\Delta}$ of the left and the right
NS supermodules is defined as a graded tensor product of representations of $\mathbb{Z}_2$-graded algebras.
The composition of tensor products of homogeneous elements is given by
$$
(A\otimes \bar A )(B\otimes \bar B)
\; = \;
(-1)^{{\rm deg}(\bar A)\cdot{\rm deg}(B)}AB\otimes \bar A \bar B\ .
$$

A nonzero element ${\chi}\in {\cal V}^{f}_\Delta $ of degree ${f}$
is called a singular vector if
it satisfies the highest weight conditions (\ref{highest}) with
$
L_0{\chi} =(\Delta +{f}){\chi}
$.
It generates its own NS supermodule ${\cal V}_{\Delta +{f}}$ which is a
submodule of ${\cal V}_{\Delta}$.

The analysis of singular vectors can be facilitated by introducing a
symmetric bilinear form $\langle
.\,,.\rangle_{c,\Delta}$ on ${\cal V}_{\Delta}$ uniquely determined  by the relations
$\langle\nu_{\Delta},\nu_{\Delta}\rangle =1$ and
$(L_{m})^{\dag}=L_{-m},(S_{k})^{\dag}=S_{-k}$.
It is block-diagonal with respect to the ${1\over 2}\mathbb Z$-grading.
We denote by $B^{\,{f}}_{c,\Delta}$ the matrix of $\langle .\,,.\rangle_{c,\Delta}$
on ${\cal V}_{\Delta}^{\,{f}}$ calculated
in the basis (\ref{basis}):
\begin{equation}
\label{matrix}
\left[ B^{\,{f}}_{c,\Delta}\right]_{KM,LN}
= \;
\langle \nu_{\Delta,KM},\nu_{\Delta,LN}\rangle_{c,\Delta}.
\end{equation}
It is nonsingular if and only if the supermodule ${\cal V}_{\Delta}$ does
not contain singular vectors of degrees ${1\over 2},1,\dots,{f}$.
The determinant of this matrix is given by the Kac theorem
\begin{equation}
\label{Kac}
\det B^{\,{f}}_{c,\Delta}
\; = \;
K_{f}\hskip -2mm
\prod\limits_{1\leqslant rs \leqslant 2{f}}
(\Delta-\Delta_{rs})^{P_{NS}({f}-{rs\over 2})}
\end{equation}
where $K_{f}$ depends only on the level $f,$ the sum $r+s$ must be even and
\begin{eqnarray}
\label{delta:rs}
\Delta_{rs}(c)
& = &
-\frac{rs-1}{4} + \frac{r^2-1}{8}\beta^2 + \frac{s^2-1}{8}\frac{1}{\beta^2}\,,
\\
\nonumber
\beta & = & \frac{1}{2\sqrt{2}}\left(\sqrt{1-\hat c} + \sqrt{9-\hat c}\right),
\hskip 10mm
\hat c = \frac23 c.
\end{eqnarray}
The multiplicity of each zero is given by
$P_{NS}({f})= \dim {\cal V}^{f}_\Delta$
and can be read off from the relation
$$
\sum\limits_{{f}=0}^\infty P_{NS}({f})q^{f}
\; = \;
\prod\limits_{n=1}^\infty {1+q^{n-{1\over 2}}\over 1-q^n}.
$$
As a function of $c$ the Kac determinant vanishes at
\begin{equation}
\label{zero:rs}
c \; = \; c_{rs}(\Delta)
\; \equiv \; \frac{3}{2} - 3\left(\beta_{rs}(\Delta) - \frac{1}{\beta_{rs}(\Delta)}\right)^2,
\end{equation}
where $1 < rs \leq 2n, \, 1 < r, \;\; r+s \in 2{\mathbb N}$, and
\[
%\begin{equation}
%\label{beta:def}
\beta^2_{rs}(\Delta)
\; = \;
\frac{1}{r^2 -1}
\left(4\Delta + rs -1 +
    \sqrt{16\Delta^2 +8(rs-1)\Delta + (r-s)^2}
\right).
%\end{equation}
\]
We shall use the following parametrization of conformal weights
which is especially useful in formulation of the fusion rules:
\begin{equation}
\label{parameterization:def}
\Delta_i
\; = \;
\frac{\hat c-1}{16} + \frac{\alpha_i^2}{8}
\; = \;
-\frac{1}{8}{\left(\beta - \frac{1}{\beta}\right)^2}+ \frac{\alpha_i^2}{8}\,.
\end{equation}
In the case of the degenerate weight one has:
\begin{equation}
\label{parameterization}
\Delta_{rs}=\frac{\hat c-1}{16} + \frac{\alpha_{rs}^2}{8}\,,
\hskip 1cm
\alpha_{rs} \; = \; r\beta - \frac{s}{\beta}\,.
\end{equation}

\section{NS chiral vertex operator}
Super-descendants $\varphi_{\Delta,\bar\Delta}(\xi,\bar\xi |z,\bar z)$
of the super-primary field $\varphi_{\Delta,\bar\Delta}(z,\bar z)=\varphi_{\Delta,\bar\Delta}(\nu,\bar\nu |z,\bar z)$ are defined by
the relations:
\begin{eqnarray*}
\varphi_{\Delta,\bar\Delta}(L_{-m}\xi,\bar\xi |z,\bar z)
&=&
\oint {dw\over 2\pi i} (w-z)^{1-m}T(w)\varphi_{\Delta,\bar\Delta}(\xi,\bar\xi |z,\bar z),
\hskip 5mm
m\in\mathbb{N},\\
\varphi_{\Delta,\bar\Delta}(S_{-k}\xi,\bar\xi |z,\bar z)
&=&
\oint {dw\over 2\pi i} (w-z)^{{1\over 2}-k}S(w)\varphi_{\Delta,\bar\Delta}(\xi,\bar\xi |z,\bar z),
\hskip 5.5mm
k\in\mathbb{N}-\textstyle{1\over 2},
\end{eqnarray*}
and by analogous formulae for the right sector. Using conformal Ward identities one can express
an arbitrary correlator  of three descendants in terms of matrix elements which can be further factorized into
the holomorphic and anti-holomorphic parts
\begin{equation}
\label{unn:rho:def}
\langle\,\xi_3,\bar\xi_3\,|\varphi_{\Delta_2,\bar\Delta_2}(\xi_2,\bar\xi_2 |z,\bar z)|\,\xi_1,\bar\xi_1\,\rangle
\; = \;
\varrho{^{\Delta_3}_\infty}{^{\Delta_2}_{\:z}}{^{\Delta_1}_{\;0}} (\xi_1,\xi_2,\xi_3)\,
\varrho{^{\bar\Delta_3}_\infty}{^{\bar\Delta_2}_{\:\bar z}}{^{\bar\Delta_1}_{\;0}}(\bar\xi_1,\bar\xi_2,\bar\xi_3),
\end{equation}
where $\xi_i\in {\cal V}_{\Delta_i}, \bar\xi_i\in {\cal V}_{\bar\Delta_i}$.
The trilinear map
$$
\varrho{^{\Delta_3}_\infty}{^{\Delta_2}_{\:z}}{^{\Delta_1}_{\;0}} :
{\cal V}_{\Delta_3}\times {\cal V}_{\Delta_2} \times {\cal
V}_{\Delta_1} \ \mapsto \ \mathbb{C}\
$$
is determined by conditions which can be easily derived by analyzing
the holomorphic (or anti-holomorphic) part of the superconformal Ward identities for the three point function
(see \cite{Teschner:2001rv}
 for an analogous  construction in the Virasoro case).
They read:
\begin{eqnarray}
\label{lcommutator}
\varrho{^{\Delta_3}_\infty}{^{\Delta_2}_{\:z}}{^{\Delta_1}_{\;0}}
(L_{-n} \xi_3,\xi_2,\xi_1)
&=&
\varrho{^{\Delta_3}_\infty}{^{\Delta_2}_{\:z}}{^{\Delta_1}_{\;0}}
( \xi_3,\xi_2,L_n\xi_1)
\\
\nonumber &+&
\sum\limits_{m=-1}^{l(n)}
 \left(
\begin{array}{c}
\scriptstyle n+1\\[-6pt]
\scriptstyle m+1
\end{array}
\right)
  z^{n-m}
  \varrho{^{\Delta_3}_\infty}{^{\Delta_2}_{\:z}}{^{\Delta_1}_{\;0}}
(\xi_3,L_m\xi_2,\xi_1),
\\[5pt]
%,,,,,,,,,,,,,,,,,,,,,,,,,,,,,,,,,,
\label{scommutator}
\varrho{^{\Delta_3}_\infty}{^{\Delta_2}_{\:z}}{^{\Delta_1}_{\;0}}
(S_{-k} \xi_3,\xi_2,\xi_1)
&=&
(-1)^{2(N(\xi_1)+N(\xi_3))}
\varrho{^{\Delta_3}_\infty}{^{\Delta_2}_{\:z}}{^{\Delta_1}_{\;0}}
( \xi_3,\xi_2,S_k\xi_1)
\\
\nonumber &+&
\sum\limits_{m=-1}^{l(k-{1\over 2})}
 \left(
\begin{array}{c}
\scriptstyle k+{1\over 2}\\[-6pt]
\scriptstyle m+1
\end{array}
\right)
  z^{k-{1\over 2} -m}
  \varrho{^{\Delta_3}_\infty}{^{\Delta_2}_{\:z}}{^{\Delta_1}_{\;0}}
(\xi_3,S_{m+{1\over 2}}\xi_2,\xi_1),
\end{eqnarray}
where
$l(m) = m$ for $ m+1 \geqslant 0$, and
$l(m)= \infty $ for $ m+1 <0$, and
\begin{eqnarray}
\label{translation}
\varrho{^{\Delta_3}_\infty}{^{\Delta_2}_{\:z}}{^{\Delta_1}_{\;0}} (
\xi_3,L_{-1}\xi_2,\xi_1)
 &=&
 \partial_z
 \varrho{^{\Delta_3}_\infty}{^{\Delta_2}_{\:z}}{^{\Delta_1}_{\;0}}
( \xi_3,\xi_2,\xi_1),
\\[5pt]
\label{L2a}
\varrho{^{\Delta_3}_\infty}{^{\Delta_2}_{\:z}}{^{\Delta_1}_{\;0}} (
\xi_3,L_{n}\xi_2,\xi_1) &=&
 \sum\limits_{m=0}^{n+1} \left(\,_{\;\;m}^{n+1}\right) (-z)^{m}
\left(\varrho{^{\Delta_3}_\infty}{^{\Delta_2}_{\:z}}{^{\Delta_1}_{\;0}}
( L_{m-n}\xi_3,\xi_2,\xi_1)\right.
\\
\nonumber &&\hspace{70pt} -\; \left.
\varrho{^{\Delta_3}_\infty}{^{\Delta_2}_{\:z}}{^{\Delta_1}_{\;0}} (
\xi_3,\xi_2,L_{n-m}\xi_1) \right), \hskip 5mm n>-1,
\end{eqnarray}
\begin{eqnarray}
\label{L2b}
\varrho{^{\Delta_3}_\infty}{^{\Delta_2}_{\:z}}{^{\Delta_1}_{\;0}} (
\xi_3,L_{-n}\xi_2,\xi_1) &=&
 \sum\limits_{m=0}^{\infty} \left(\,_{\;\;n-2}^{n-2+m}\right)
z^{m}
\varrho{^{\Delta_3}_\infty}{^{\Delta_2}_{\:z}}{^{\Delta_1}_{\;0}}
(L_{n+m} \xi_3,\xi_2,\xi_1)
\\
\nonumber && \hspace{-40pt} +\;(-1)^n \sum\limits_{m=0}^{\infty}
\left(\,_{\;\;n-2}^{n-2+m}\right) z^{-n+1-m}
\varrho{^{\Delta_3}_\infty}{^{\Delta_2}_{\:z}}{^{\Delta_1}_{\;0}} (
\xi_3,\xi_2, L_{m-1}\xi_1), \hskip 5mm n>1,
\\[15pt]
%,,,,,,,,,,,,,,,,,,,,,,,,,,,,,,,,,,,,,,,,,,,,,,,,,,
\label{S2a}
\varrho{^{\Delta_3}_\infty}{^{\Delta_2}_{\:z}}{^{\Delta_1}_{\;0}} (
\xi_3,S_{k}\xi_2,\xi_1) &=&
 \sum\limits_{m=0}^{k+{1\over 2}}
 \left(
\begin{array}{c}
\scriptstyle k+{1\over 2}\\[-6pt]
\scriptstyle m
\end{array}
\right)
 (-z)^{m}
\left(\varrho{^{\Delta_3}_\infty}{^{\Delta_2}_{\:z}}{^{\Delta_1}_{\;0}}
( S_{m-k}\xi_3,\xi_2,\xi_1)\right.
\\
\nonumber &&\hspace{20pt} +\ (-1)^{2(N(\xi_1)+N(\xi_3))}\; \left.
\varrho{^{\Delta_3}_\infty}{^{\Delta_2}_{\:z}}{^{\Delta_1}_{\;0}} (
\xi_3,\xi_2,S_{k-m}\xi_1) \right),
\hskip 5mm k\geqslant-\scriptstyle {1\over 2},
%,,,,,,,,,,,,,,,,,,,,,,,,,,,,,,,,,,,
\\[10pt]
\label{S2b}
\varrho{^{\Delta_3}_\infty}{^{\Delta_2}_{\:z}}{^{\Delta_1}_{\;0}} (
\xi_3,S_{-k}\xi_2,\xi_1) &=&
 \sum\limits_{m=0}^{\infty}
 \left(
\begin{array}{c}
\scriptstyle k-{3\over 2}+m\\[-6pt]
\scriptstyle m
\end{array}
\right)
z^{m}
\varrho{^{\Delta_3}_\infty}{^{\Delta_2}_{\:z}}{^{\Delta_1}_{\;0}}
(S_{k+m} \xi_3,\xi_2,\xi_1)
\\
\nonumber && \hspace{-95pt} +\;
(-1)^{2(N(\xi_1)+N(\xi_3))+k+{1\over 2}}
\sum\limits_{m=0}^{\infty}
 \left(
\begin{array}{c}
\scriptstyle k-{3\over 2}+m\\[-6pt]
\scriptstyle m
\end{array}
\right)
 z^{-k-m+{1\over 2}}
\varrho{^{\Delta_3}_\infty}{^{\Delta_2}_{\:z}}{^{\Delta_1}_{\;0}} (
\xi_3,\xi_2, S_{m-{1\over 2}}\xi_1), \hskip 5mm k>\scriptstyle {1\over 2}.
\end{eqnarray}
The form
$\displaystyle
\varrho{^{\Delta_3}_\infty}{^{\Delta_2}_{\:z}}{^{\Delta_1}_{\;0}}$ is
almost completely determined by the properties above. In particular, for
$L_0$-eingenstates, $ L_0\,\xi_i  = \Delta_i(\xi_i)\xi_i,\;i=1,2,3,$ one has:
\begin{equation}
\label{z:dep}
\varrho{^{\Delta_3}_\infty}{^{\Delta_2}_{\:z}}{^{\Delta_1}_{\;0}}
(\xi_3,\xi_2,\xi_1) \; = \; z^{\Delta_3(\xi_3)- \Delta_2(\xi_2)-\Delta_1(\xi_1)}\
\varrho{^{\Delta_3}_\infty}{^{\Delta_2}_{\:1}}{^{\Delta_1}_{\;0}}
(\xi_3,\xi_2,\xi_1).
\end{equation}
However,
in contrast to the case of the Virasoro algebra, the form
$\varrho^{\Delta_3\ \Delta_2 \ \Delta_1}_{\infty \ \ 1 \ \ \ 0}$
is determined up to two,  instead of one, independent constants:
\begin{eqnarray*}
A&=& \varrho^{\Delta_3\ \Delta_2 \ \Delta_1}_{\infty \ \ 1 \ \ \ 0}
(\nu_3, \nu_2 , \nu_1 ),
\\
B&=& \varrho^{\Delta_3\ \Delta_2 \ \Delta_1}_{\infty \ \ 1 \ \ \ 0}
(\nu_3, *\nu_2 , \nu_1 )
    = \varrho^{\Delta_3\ \Delta_2 \ \Delta_1}_{\infty \ \ 1 \ \ \ 0} (*\nu_3, \nu_2 , \nu_1 )
    = \varrho^{\Delta_3\ \Delta_2 \ \Delta_1}_{\infty \ \ 1 \ \ \ 0} (\nu_3, \nu_2 , *\nu_1 ),
\end{eqnarray*}
where $\nu_i$ is the highest weight state in ${\cal V}_{\Delta_i}$
($i=1,2,3$) and
\begin{equation}
\label{star}
*\!\nu_i
\; \equiv \;
S_{-{1\over 2}}\nu_i.
\end{equation}
For this reason the NS superconformal theory requires
two independent structure constants (and in consequence 8  conformal blocks).
Indeed, taking into account that the superprimary fields are even
with respect to the common parity operator $(-1)^F\otimes (-1)^{\bar F}$
one can show that the constants $A_L,B_L$ of the left sector and the constants
$A_R,B_R$ of the right sector always show up in the combinations
$A_LA_R$, $B_LB_R$ which are just  the structure constants (\ref{constant1}), (\ref{constant2}):
\begin{eqnarray*}
A_LA_R& = &C_{321}=  \Big\langle\varphi_3(\infty,\infty)\varphi_2(1,1)\varphi_1(0,0)\Big\rangle,
\\[6pt]
B_LB_R& = & \widetilde C_{321}= \Big\langle\varphi_3(\infty,\infty)\widetilde\varphi_2(1,1)\varphi_1(0,0)\Big\rangle.
\end{eqnarray*}
As it is more convenient to keep in the correlation functions an explicit $C_{321}$ and $\widetilde C_{321}$ dependence
we shall work in the following mostly with the normalized form.
In order to avoid  confusion with the unnormalized form $\varrho^{\Delta_3\ \Delta_2 \ \Delta_1}_{\infty \ \ z \ \ \ 0}$
we shall denote the normalized one by
$\rho^{\Delta_3\ \Delta_2 \ \Delta_1}_{\infty \ \ z \ \ \ 0}.$ Thus, by definition
\[
\rho^{\Delta_3\ \Delta_2 \ \Delta_1}_{\infty \ \ z \ \ \ 0}(\nu_3,\nu_2,\nu_1)
\; = \;
\rho^{\Delta_3\ \Delta_2 \ \Delta_1}_{\infty \ \ z \ \ \ 0}(\nu_3,*\nu_2,\nu_1)
\; = \; 1.
\]

For each $\xi_2 \in {\cal V}_{\Delta_2}$ we define the (generalized)  chiral vertex $V(\xi_2|z)$
\cite{Moore:1988qv}
as a linear map
$$
V(\xi_2|z):\; {\cal V}_{\Delta_1} \; \mapsto \; {\cal V}_{\Delta_3}\,,
$$
such that
$$
\langle\,\xi_3\,|V(\xi_2|z)|\,\xi_1\,\rangle
\; = \;
\rho{^{\Delta_3}_\infty}{^{\Delta_2}_{\:z}}{^{\Delta_1}_{\;0}}
(\xi_3,\xi_2,\xi_1)
$$
for all $\xi_3 \in {\cal V}_{\Delta_3}, \xi_1 \in {\cal V}_{\Delta_1}$.
Note that $V(\xi_2|z)$ does not have definite parity with respect to
the left fermionic number $(-1)^F$ and can be  decomposed into its
even (parity preserving) and odd (parity reversing) part:
$$
V(\xi_2|z)=V^{\rm \scriptscriptstyle even}(\xi_2|z)+V^{\rm \scriptscriptstyle odd}(\xi_2|z).
$$
We shall need  two special cases of these operators:
$V(\nu_2|z)$ and $V(*\nu_2|z)$. Formulae (\ref{lcommutator}), (\ref{scommutator})
and (\ref{translation}) imply  the following relations:
\begin{eqnarray}
\label{Lkom}
\nonumber
  \left[ \mathrm{L}_m, V(\nu_2 | z)  \right]
  &=& z^m \left( z \partial_z + (m+1) \Delta_2 \right)V(\nu_2 | z),
\\[4pt]
\nonumber
 \left[ \mathrm{L}_m, V(*\nu_2 | z)  \right]
    &=& z^m \left( z \partial_z + (m+1) \left(\Delta_2 + \textstyle \frac{1}{2}\right)  \right)
    V(*\nu_2 | z),
\\[10pt]
 \label{Skom1}
 \nonumber
  \left[ \mathrm{S}_k, V^{\rm \scriptscriptstyle even}(\nu_2 | z)  \right]
  &=& z^{k+ \frac{1}{2}}\ V^{\rm \scriptscriptstyle odd}(*\nu_2 | z),
\\[4pt]
\left\{ \mathrm{S}_k, V^{\rm \scriptscriptstyle odd}(\nu_2 | z)  \right\}
  &=& z^{k+ \frac{1}{2}}\ V^{\rm \scriptscriptstyle even}(*\nu_2 | z),
\\[10pt]
\label{Skom2}
\nonumber
 \left[ \mathrm{S}_k, V^{\rm \scriptscriptstyle even}(*\nu_2 | z)  \right]
    &=& z^{k - \frac{1}{2}} \left( z \partial_z + \Delta_2 (2k+1) \right)
   \ V^{\rm \scriptscriptstyle odd}(\nu_2 | z),
\\[4pt]
\nonumber
 \left\{ \mathrm{S}_k, V^{\rm \scriptscriptstyle odd}(*\nu_2 | z)  \right\}
    &=& z^{k - \frac{1}{2}} \left( z \partial_z + \Delta_2 (2k+1) \right)
   \ V^{\rm \scriptscriptstyle even}(\nu_2 | z).
\end{eqnarray}
Using them one gets:
{\small
\begin{eqnarray}
\label{matrixele1}
\rho^{\Delta_3\ \Delta_2 \ \Delta_1}_{\infty \ \ z \ \ \ 0}
( \nu_{3,KM}, \nu_2 , \nu_1 )
&=&  z^{\Delta_3 + |K| + |M| - \Delta_2 -  \Delta_1}
\left\{
\begin{array}{lcl}
 \eta^{\rm{o}}_{\Delta_3+|M|} \left[^{\Delta_2}_{\Delta_1} \right]_K
\gamma_{\Delta_3} \left[^{\Delta_2}_{\Delta_1} \right]_M,
 && %|K| \in \mathbb{N}
\\
\\
 \eta^{\rm{e}}_{\Delta_3+|M|} \left[^{\Delta_2}_{\Delta_1} \right]_K
\gamma_{\Delta_3} \left[^{\Delta_2 + \frac{1}{2}}_{\hskip 8pt \Delta_1} \right]_M,
&& %|K| \in \mathbb{N}- \frac{1}{2}
\end{array}
\right.
\\[10pt]
\label{matrixele2}
\rho^{\Delta_3\ \Delta_2 \ \Delta_1}_{\infty \ \ z \ \ \ 0}
(\nu_{3,KM} , *\nu_2 , \nu_1 )
&=&  z^{\Delta_3 + |K|+|M| - \Delta_2 -  \Delta_1 - \frac{1}{2}}
\left\{
\begin{array}{lcl}
 \eta^{\rm{e}}_{\Delta_3+|M|} \left[^{\Delta_2}_{\Delta_1} \right]_K
\gamma_{\Delta_3} \left[^{\Delta_2 + \frac{1}{2}}_{\hskip 8pt \Delta_1} \right]_M,
 %&&  I \in \mathbb{N}
\\
\\
 \eta^{\rm{o}}_{\Delta_3+|M|} \left[^{\Delta_2}_{\Delta_1} \right]_K
\gamma_{\Delta_3} \left[^{\Delta_2}_{\Delta_1} \right]_M,
% &&  I \in \mathbb{N}+ \frac{1}{2}
\end{array}
\right.
  \end{eqnarray}
}

\noindent
where the upper lines correspond to $|K| \in \mathbb{N} $,
 the lower lines to $|K| \in \mathbb{N}- \frac{1}{2} $, and
{\small
\begin{eqnarray*}
\gamma_\Delta\!\left[^{\Delta_2}_{\Delta_1}\right]_{M}
& \stackrel{\rm def}{=} &
\left(\Delta - \Delta_1 + m_1\Delta_2\right)
\left(\Delta - \Delta_1 + m_2\Delta_2+ m_1\right)
\cdots
\left(\Delta - \Delta_1 + m_j\Delta_2+ \sum\limits_{l=1}^{j-1}m_l\right),
\\[6pt]
\eta^{\rm o}_\Delta\!\left[^{\Delta_2}_{\Delta_1}\right]_{K}
& \stackrel{\rm def}{=} &
\left(\Delta - \Delta_1 + 2k_1\Delta_2\right)
\left(\Delta - \Delta_1 + 2k_3\Delta_2 +k_1 + k_2\right)\ldots
\left(\Delta - \Delta_1 + 2k_{p}\Delta_2+ \sum\limits_{l=1}^{p-1}k_l\right),
\\[6pt]
\eta^{\rm e}_\Delta\!\left[^{\Delta_2}_{\Delta_1}\right]_{K}
& \!\stackrel{\rm def}{=} &
\left(\Delta - \Delta_1 + 2k_2\Delta_2 + k_1\right)
\left(\Delta - \Delta_1 + 2k_4\Delta_2 + \sum\limits_{l=1}^{3}k_l\right)
\!..\!
\left(\Delta - \Delta_1 + 2k_{p'}\Delta_2 + \!\sum\limits_{l=1}^{p'-1}k_l\right)\!,
\end{eqnarray*}
}

\noindent
where $p$ is the largest odd number not grater than $i$, and
$p'$ is the largest even number not grater than $i$.

For the matrix elements
$ \bra{\nu_3 }  V(\underline{\hspace{4pt}}\,\nu_2 | z)  \ket{ \nu_{1,KM}}$
one gets:
\begin{equation}
\label{reflection1}
\begin{array}{lcl}
\rho^{\Delta_3\ \Delta_2 \ \Delta_1}_{\infty \ \ 1 \ \ \ 0}
(\nu_3,
\nu_2 , \nu_{1,KM} )
 & = &
\rho^{\Delta_1\ \Delta_2 \ \Delta_3}_{\infty \ \ 1 \ \ \ 0}
(\nu_{1,KM}, \nu_3 , \nu_4
),
\\[6pt]
\rho^{\Delta_3\ \Delta_2 \ \Delta_1}_{\infty \ \ 1 \ \ \ 0}
(\nu_3, *\nu_2 , \nu_{1,KM} )
& = &
\rho^{\Delta_1\ \Delta_2 \ \Delta_3}_{\infty \ \ 1 \ \ \ 0}
(\nu_{1,KM}, *\nu_2 , \nu_3 ),
\end{array}
\end{equation}
for $|K| \in {\mathbb N}\cup \{0\}$, and
\begin{equation}
\label{reflection2}
\begin{array}{lcl}
\rho^{\Delta_3\ \Delta_2 \ \Delta_1}_{\infty \ \ 1 \ \ \ 0}
(\nu_3, \nu_2 , \nu_{1,KM} )
& = &
\rho^{\Delta_1\ \Delta_2 \ \Delta_3}_{\infty \ \ 1 \ \ \ 0}
(\nu_{1,KM}, \nu_2 , \nu_3 ),
\\[6pt]
\rho^{\Delta_3\ \Delta_2 \ \Delta_1}_{\infty \ \ 1 \ \ \ 0}
(\nu_3, *\nu_2 , \nu_{1,KM} )
& = &
-\ \rho^{\Delta_1\ \Delta_2 \ \Delta_3}_{\infty \ \ 1 \ \ \ 0}
(\nu_{1,KM}, *\nu_2 , \nu_3 ),
\end{array}
\end{equation}
for $|K| \in {\mathbb N}-\frac12$.

Using formulae (\ref{matrixele1}) and (\ref{matrixele2}) one can show an important factorization
property of the  form $\varrho^{\Delta_3\ \Delta_2 \ \Delta_1}_{\infty \ \ 1 \ \ \ 0}$.
For an arbitrary $L_0$-eingenstate $ L_0\,\xi_3 = (\Delta_3+{f})\,\xi_3$ one has:
\begin{eqnarray}
\nonumber
&& \hspace{-30pt}
\varrho^{\Delta_3\ \Delta_2 \ \Delta_1}_{\infty \ \ z \ \ \ 0} (\mathrm{S}_{-K} \mathrm{L}_{-M}\, \xi_3 , \nu_2 , \nu_1 )
\; = \;
z^{\Delta_3 + {f} +|K|+|M| - \Delta_2 -  \Delta_1} \times
\\[6pt]
\nonumber
&& \hspace{30pt}
\rho^{\Delta_3+ {f}\ \Delta_2 \ \Delta_1}_{\ \ \infty   \ \ \ \, 1 \ \ \ 0} (\mathrm{S}_{-K} \mathrm{L}_{-M} \nu_{\Delta_3+{f}}, \nu_2 , \nu_1 ) \
\times
\left\{
\begin{array}{lcl}
\varrho^{\Delta_3\ \Delta_2 \ \Delta_1}_{\infty \ \ 1 \ \ \ 0}(\xi_3, \nu_2 , \nu_1 ),
\\[8pt]
%\rho^{\Delta_3+ {f}\ \Delta_2 \ \Delta_1}_{\ \ \infty   \ \ \ \, 1 \ \ \ 0} (\mathrm{S}_{-K} \mathrm{L}_{-M} \nu_{\Delta_3+{f}},  \nu_2 , \nu_1 ) \
\varrho^{\Delta_3\ \Delta_2 \ \Delta_1}_{\infty \ \ 1 \ \ \ 0}
(\xi_3, *\nu_2 , \nu_1 ),
\end{array}
\right.
\\[10pt]
\label{factorization}
&&  \hspace{-30pt}
\varrho^{\Delta_3\ \Delta_2 \ \Delta_1}_{\infty \ \ z \ \ \ 0}
    (\mathrm{S}_{-K} \mathrm{L}_{-M}\, \xi_3 , *\nu_2 , \nu_1 )
=  z^{\Delta_3 + {f} +|K|+|M|- \Delta_2 -  \Delta_1 -
\frac{1}{2}} \times
\\[8pt]
\nonumber
&& \hspace{30pt}
\rho^{\Delta_3+ {f}\ \Delta_2 \ \Delta_1}_{\ \ \infty   \ \ \ \, 1 \ \ \ 0} (\mathrm{S}_{-K} \mathrm{L}_{-M} \nu_{\Delta_3+{f}}, *\nu_2 , \nu_1 ) \
\times
\left\{
\begin{array}{lcl}
\varrho^{\Delta_3\ \Delta_2 \ \Delta_1}_{\infty \ \ 1 \ \ \ 0}(\xi_3, *\nu_2 , \nu_1 ),
\\[10pt]
\varrho^{\Delta_3\ \Delta_2 \ \Delta_1}_{\infty \ \ 1 \ \ \ 0}(\xi_3, \nu_2 , \nu_1 ),
\end{array}
\right.
  \end{eqnarray}
where the upper lines correspond  to $|K| \in \mathbb{N}\cup\{0\} $, and the lower lines to  $|K|
\in \mathbb{N}-\frac{1}{2} $.

\section{NS superconformal blocks}
\label{definitions}
W shall define four  types of NS superconformal blocks directly in
terms of the form $\rho$. For each type there is one even,
\begin{eqnarray*}
\mathcal{F}^1_{\Delta}
\left[^{\underline{\hspace{3pt}}\,\Delta_3
\;\underline{\hspace{3pt}}\,\Delta_2}_{\hspace{3pt}\,\Delta_4
\;\hspace{3pt}\, \Delta_1} \right]
(z)
    &=&
z^{\Delta - \underline{\hspace{3pt}}\,\Delta_2 - \Delta_1} \left( 1 +
\sum_{m\in \mathbb{N}} z^m
    F^m_{c, \Delta}
    \left[^{\underline{\hspace{3pt}}\,\Delta_3
\;\underline{\hspace{3pt}}\,\Delta_2}_{\hspace{3pt}\,\Delta_4
\;\hspace{3pt}\, \Delta_1} \right]
     \right),
\end{eqnarray*}
and one odd,
\begin{eqnarray*}
\mathcal{F}^{\frac{1}{2}}_{\Delta}
\left[^{\underline{\hspace{3pt}}\,\Delta_3
\;\underline{\hspace{3pt}}\,\Delta_2}_{\hspace{3pt}\,\Delta_4
\;\hspace{3pt}\, \Delta_1} \right]
(z)
    &=&
z^{\Delta - \underline{\hspace{3pt}}\,\Delta_2 - \Delta_1 }
\sum_{k\in \mathbb{N}- \frac{1}{2}} z^k
        F^k_{c, \Delta}
        \left[^{\underline{\hspace{3pt}}\,\Delta_3
\;\underline{\hspace{3pt}}\,\Delta_2}_{\hspace{3pt}\,\Delta_4
\;\hspace{3pt}\, \Delta_1} \right],
\end{eqnarray*}
conformal block.

The coefficients are defined by:
\begin{eqnarray}
\label{block:definition}
&&
F^{f}_{c, \Delta}
\left[^{\underline{\hspace{3pt}}\,\Delta_3
\;\underline{\hspace{3pt}}\,\Delta_2}_{\hspace{3pt}\,\Delta_4
\;\hspace{3pt}\, \Delta_1} \right] \; =
\\
\nonumber
&& =
\hspace*{-20pt}
\begin{array}[t]{c}
{\displaystyle\sum} \\[2pt]
{\scriptstyle
|K|+|M| = |L|+|N| = f
}
\end{array}
\hspace*{-20pt}
\rho^{\Delta_4\ \Delta_3 \ \Delta}_{\infty \ \ 1 \ \ \ 0} (\nu_4, \underline{\hspace{4pt}}\,\nu_3 , \nu_{\Delta,KM} )
\ \left[B^{f}_{c, \Delta}\right]^{KM,LN}  \rho^{\Delta\ \Delta_2 \ \Delta_1}_{\infty \ \ 1 \ \ \ 0}
   (\nu_{\Delta,LN},  \underline{\hspace{4pt}}\,\nu_2 , \nu_1 ),
\end{eqnarray}
where {\small $\left[B^{f}_{c, \Delta}\right]^{KM,LN}$} is the matrix inverse to the matrix (\ref{matrix}),
$\underline{\hspace{4pt}}\,\Delta_i$ and $\underline{\hspace{4pt}}\,\nu_i$ stand for $\Delta_i$ or $*\Delta_i$,
and $\nu_i$ or $*\nu_i$, respectively, and
$
z^{\Delta -*\Delta_2 - \Delta_1} =z^{\Delta - \Delta_2 - \Delta_1-{1\over 2}}\ .
$
For instance:
\begin{eqnarray*}
\mathcal{F}^1_{\Delta}\!
\left[^{\Delta_3
\;*\Delta_2}_{\Delta_4
\;\hspace{4.5pt}\Delta_1} \right]
(z)
    &=&
z^{\Delta - \Delta_2 - \Delta_1-{1\over 2}} \left( 1 +
\sum_{m\in \mathbb{N}} z^m
    F^m_{c, \Delta}
    \left[^{\Delta_3
\;*\Delta_2}_{\Delta_4
\;\hspace{4.5pt}\Delta_1} \right]
     \right),
\\[4pt]
%,,,,,,,,,,,,,,,,,,,,,,,,,
F^m_{c, \Delta} \left[^{\Delta_3
\;*\Delta_2}_{\Delta_4
\;\hspace{4.5pt}\Delta_1} \right]
    &=&
     \sum
    \rho^{\Delta_4\ \Delta_3 \ \Delta}_{\infty \ \ 1 \ \ \ 0} (\nu_4, \nu_3 , \nu_{\Delta,KM} )
   \ \left[B^{\ m}_{c, \Delta}\right]^{KM,LN} \ \rho^{\Delta\ \Delta_2 \ \Delta_1}_{\infty \ \ 1 \ \ \ 0}
   (\nu_{\Delta,LN}, *\nu_2 , \nu_1 ).
\end{eqnarray*}
Let us note that  the definition above is  independent of the choice of
basis in ${\cal V}_\Delta^{f}$.

The formulae (\ref{reflection1}), (\ref{reflection2}) imply simple relations:
\begin{eqnarray*}
\mathcal{F}^1_{\Delta}\!
\left[^{\Delta_3\;*\Delta_2}_{\Delta_4\;\hspace{4.5pt}\Delta_1} \right]\!(z)
&=&
z^{ -\Delta_2 - \Delta_1-{1\over 2}+ \Delta_4 + \Delta_3}\
\mathcal{F}^1_{\Delta}\!
\left[^{*\Delta_2 \;\Delta_3}_{\hspace{4.5pt}\Delta_1\;\Delta_4} \right]\!(z),
\\[10pt]
\mathcal{F}^{1\over 2}_{\Delta}\!
\left[^{\Delta_3\;*\Delta_2}_{\Delta_4\;\hspace{4.5pt}\Delta_1} \right]\!(z)
&=&
-z^{ -\Delta_2 - \Delta_1-{1\over 2}+ \Delta_4 + \Delta_3}\
\mathcal{F}^{1\over 2}_{\Delta}\!
\left[^{*\Delta_2 \;\Delta_3}_{
        \hspace{4.5pt}\Delta_1\;\Delta_4} \right]\!(z),
\end{eqnarray*}
reducing the number of blocks to 6 independent functions.

The primary fields are expressed in terms of chiral vertex operators as follows:
\begin{eqnarray}
\label{phiv}\nonumber
\phi_2(z, \bar{z}) &=&  \bigoplus_{\Delta_3, \Delta_1}
\Bigg(
    C_{321} V^{\rm \scriptscriptstyle even}(\nu_2|z)\otimes
    V^{\rm \scriptscriptstyle even}(\bar \nu_2|\bar z)
    -\tilde C_{321} V^{\rm \scriptscriptstyle odd}(\nu_2|z)\otimes
    V^{\rm \scriptscriptstyle odd}(\bar \nu_2|\bar z)
\Bigg),
\\[10pt]
\label{psiv}\nonumber
\psi_2(z, \bar{z}) &=&  \bigoplus_{\Delta_3, \Delta_1}
\Bigg(
    C_{321} V^{\rm \scriptscriptstyle odd}(*\nu_2|z)\otimes
    V^{\rm \scriptscriptstyle even}(\bar \nu_2|\bar z)
    -\tilde C_{321} V^{\rm \scriptscriptstyle even}(*\nu_2|z)\otimes
    V^{\rm \scriptscriptstyle odd}(\bar \nu_2|\bar z)
\Bigg),
\\[10pt]
\label{barpsiv}\nonumber
\bar{\psi}_2(z, \bar{z}) &=&  \bigoplus_{\Delta_3, \Delta_1}
\Bigg(
    C_{321} V^{\rm \scriptscriptstyle even}(\nu_2|z)\otimes
    V^{\rm \scriptscriptstyle odd}(*\bar \nu_2|\bar z)
   +\tilde C_{321} V^{\rm \scriptscriptstyle odd}(\nu_2|z)\otimes
    V^{\rm \scriptscriptstyle even}(*\bar \nu_2|\bar z)
\Bigg),
\\[10pt]
\label{tildephiv}\nonumber
\tilde{\phi}_2(z, \bar{z})  &=&  \bigoplus_{\Delta_3, \Delta_1}
\Bigg(
    C_{321} V^{\rm \scriptscriptstyle odd}(*\nu_2|z)\otimes
    V^{\rm \scriptscriptstyle odd}(*\bar \nu_2|\bar z)
    +\tilde C_{321} V^{\rm \scriptscriptstyle even}(*\nu_2|z)\otimes
    V^{\rm \scriptscriptstyle even}(*\bar \nu_2|\bar z)
\Bigg).
\end{eqnarray}

Using this representation  and factorization in the (not orthogonal)
basis (\ref{basis}) one gets the following expressions for all basic 4-point functions
(for simplicity we write the expressions in the diagonal case $\Delta_i=\bar\Delta_i$):
\newpage
\begin{eqnarray*}
\bra{\Delta_4} \phi_3(1,1) \phi_2(z, \bar{z}) \ket{\Delta_1}
&=&
\\
&&\hspace{-80pt}\sum_p \Big(
         C_{43p} C_{p21} \left|\mathcal{F}_{\Delta_p}^{1}\!
         \left[_{\Delta_4 \ \Delta_1}^{\Delta_3 \ \Delta_2} \right]\!(z)\right|^2
         -
         \tilde{C}_{43p} \tilde{C}_{p21}  \left| \mathcal{F}_{\Delta_p}^{\frac{1}{2}}\!
         \left[_{\Delta_4 \ \Delta_1}^{\Delta_3 \ \Delta_2} \right]\!(z) \right|^2
         \Big),
\\[8pt]
%222222222222222222222222222222222
\bra{\Delta_4} \phi_3(1,1) \tilde \phi_2(z, \bar{z}) \ket{\Delta_1}
&=&
\\
&&\hspace{-80pt}\sum_p \Big(
        C_{43p} \tilde   C_{p21} \left|\mathcal{F}_{\Delta_p}^{1}\!
         \left[^{\Delta_3
\;*\Delta_2}_{\Delta_4
\;\hspace{4.5pt}\Delta_1} \right]
         (z)\right|^2
         +
          \tilde{C}_{43p} {C}_{p21}  \left| \mathcal{F}_{\Delta_p}^{1\over 2}\!
\left[^{\Delta_3
\;*\Delta_2}_{\Delta_4
\;\hspace{4.5pt}\Delta_1} \right]
         (z) \right|^2
         \Big),
\\[9pt]
%33333333333333333333333333333333333333
\bra{\Delta_4} \tilde \phi_3(1,1) \phi_2(z, \bar{z}) \ket{\Delta_1}
&=&
\\
&&\hspace{-80pt}\sum_p \Big(
         \tilde C_{43p} C_{p21} \left|\mathcal{F}_{\Delta_p}^{1}\!
         \left[^{*\Delta_3
\;\Delta_2}_{\hspace{4.5pt}\Delta_4
\;\Delta_1} \right]
         (z)\right|^2
         +
          {C}_{43p} \tilde{C}_{p21}  \left| \mathcal{F}_{\Delta_p}^{1\over 2}\!
\left[^{*\Delta_3
\;\Delta_2}_{\hspace{4.5pt}\Delta_4
\;\Delta_1} \right]
         (z) \right|^2
         \Big),
\\[9pt]
%44444444444444444444444444444444444
\bra{\Delta_4} \tilde \phi_3(1,1) \tilde \phi_2(z, \bar{z}) \ket{\Delta_1}
&=&
\\
&&\hspace{-80pt}\sum_p \Big(
         \tilde C_{43p} \tilde C_{p21} \left|\mathcal{F}_{\Delta_p}^{1}\!
         \left[^{*\Delta_3
\;*\Delta_2}_{\hspace{4.5pt}\Delta_4
\;\hspace{4.5pt}\Delta_1} \right]
         (z)\right|^2
         -
         {C}_{43p} {C}_{p21}  \left| \mathcal{F}_{\Delta_p}^{1\over 2}\!
\left[^{*\Delta_3
\;*\Delta_2}_{\hspace{4.5pt}\Delta_4
\;\hspace{4.5pt}\Delta_1} \right]
         (z) \right|^2
         \Big),
\\[9pt]
%555555555555555555555555555555555555
\bra{\Delta_4}  \psi_3(1,1) \psi_2(z, \bar{z}) \ket{\Delta_1}
&=&
\\
&&\hspace{-80pt}\sum_p
        \left(
         \tilde C_{43p} \tilde C_{p21}
         \mathcal{F}_{\Delta_p}^{1}\!
        \left[^{*\Delta_3
\;*\Delta_2}_{\hspace{4.5pt}\Delta_4
\;\hspace{4.5pt}\Delta_1} \right]\!(z)
         \mathcal{F}_{\Delta_p}^{\frac{1}{2}}\!
        \left[_{\Delta_4 \  \Delta_1}^{ \Delta_3 \ \Delta_2} \right](\bar{z})
        \right.
\\
&&\hspace{-35pt}         + \left. {C}_{43p} {C}_{p21}
\mathcal{F}_{\Delta_p}^{\frac{1}{2}}\!
         \left[^{*\Delta_3
\;*\Delta_2}_{\hspace{4.5pt}\Delta_4
\;\hspace{4.5pt}\Delta_1} \right]\!(z)
        \mathcal{F}_{\Delta_p}^{1}\!
         \left[_{ \Delta_4 \ \Delta_1}^{\Delta_3 \  \Delta_2} \right](\bar{z})
        \right),
\\[6pt]
%666666666666666666666666666666666666666666
\bra{\Delta_4}  \bar \psi_3(1,1) \bar \psi_2(z, \bar{z}) \ket{\Delta_1}
&=&
\\
&&\hspace{-80pt}\sum_p
        \left(
         C_{43p} C_{p21}
        \mathcal{F}_{\Delta_p}^{1}\!
         \left[_{ \Delta_4 \ \Delta_1}^{\Delta_3 \  \Delta_2} \right]\!(z)
          \mathcal{F}_{\Delta_p}^{\frac{1}{2}}\!
         \left[^{*\Delta_3
\;*\Delta_2}_{\hspace{4.5pt}\Delta_4
\;\hspace{4.5pt}\Delta_1} \right]\!(\bar{z})
        \right.
\\
&&\hspace{-35pt}         + \left.  \tilde{C}_{43p} \tilde{C}_{p21}
 \mathcal{F}_{\Delta_p}^{\frac{1}{2}}\!
        \left[_{\Delta_4 \  \Delta_1}^{ \Delta_3 \ \Delta_2} \right]
            (z)
\mathcal{F}_{\Delta_p}^{1}\!
        \left[^{*\Delta_3
\;*\Delta_2}_{\hspace{4.5pt}\Delta_4
\;\hspace{4.5pt}\Delta_1} \right]
        (\bar{z})
        \right),
\\[6pt]
%77777777777777777777777777777777777777777777
\bra{\Delta_4}   \psi_3(1,1) \bar \psi_2(z, \bar{z}) \ket{\Delta_1}
&=&
\\
&&\hspace{-80pt}\sum_p
        \left(-
         \tilde C_{43p} C_{p21}
 \mathcal{F}_{\Delta_p}^{1}\!
        \left[^{*\Delta_3
\;\Delta_2}_{\hspace{4.5pt}\Delta_4
\;\Delta_1} \right]
            (z)
\mathcal{F}_{\Delta_p}^{1\over 2}\!
        \left[^{\Delta_3
\;*\Delta_2}_{\Delta_4
\;\hspace{4.5pt}\Delta_1} \right]
        (\bar{z})
        \right.
\\
&&\hspace{-35pt}         + \left.  {C}_{43p} \tilde{C}_{p21}
\mathcal{F}_{\Delta_p}^{1\over 2}\!
        \left[^{*\Delta_3
\;\Delta_2}_{\hspace{4.5pt}\Delta_4
\;\Delta_1} \right]
         (z)
          \mathcal{F}_{\Delta_p}^{1}\!
         \left[^{\Delta_3
\;*\Delta_2}_{\Delta_4
\;\hspace{4.5pt}\Delta_1} \right]\!(\bar{z})
        \right),
\\[6pt]
%8888888888888888888888888888888888888
\bra{\Delta_4}  \bar \psi_3(1,1) \psi_2(z, \bar{z}) \ket{\Delta_1}
&=&
\\
&&\hspace{-80pt}\sum_p
        \left(
         - C_{43p}\tilde C_{p21}
        \mathcal{F}_{\Delta_p}^{1}\!
         \left[^{\Delta_3
\;*\Delta_2}_{\Delta_4
\;\hspace{4.5pt}\Delta_1} \right]
         (z)
          \mathcal{F}_{\Delta_p}^{\frac{1}{2}}\!
         \left[^{*\Delta_3
\;\Delta_2}_{\hspace{4.5pt}\Delta_4
\;\Delta_1} \right]\!(\bar{z})
        \right.
\\
&&\hspace{-35pt}         + \left. \tilde{C}_{43p} {C}_{p21}
 \mathcal{F}_{\Delta_p}^{\frac{1}{2}}\!
        \left[^{\Delta_3
\;*\Delta_2}_{\Delta_4
\;\hspace{4.5pt}\Delta_1} \right]
            (z)
\mathcal{F}_{\Delta_p}^{1}\!
        \left[^{*\Delta_3
\;\Delta_2}_{\hspace{4.5pt}\Delta_4
\;\Delta_1} \right]
        (\bar{z})
        \right).
\vspace*{-10pt}
\end{eqnarray*}

\newpage
\section{Recurrence relations }
%\vspace*{-12pt}
It follows from the definition of the blocks' coefficients
$
F^{f}_{c, \Delta}\!
\left[^{\underline{\hspace{3pt}}\,\Delta_3
\;\underline{\hspace{3pt}}\,\Delta_2}_{\hspace{3pt}\,\Delta_4
\;\hspace{3pt}\, \Delta_1} \right]
$
 that they are polynomials in the external weights $\Delta_i$,
and rational functions of the intermediate weight $\Delta$ and the central charge $c.$ Consequently, they
can be expressed either as a sum over the poles in~$\Delta:$
\begin{equation}
\label{first:expansion:Delta}
F^{f}_{c, \Delta}\!
\left[^{\underline{\hspace{3pt}}\,\Delta_3
\;\underline{\hspace{3pt}}\,\Delta_2}_{\hspace{3pt}\,\Delta_4
\;\hspace{3pt}\, \Delta_1} \right]
\; = \;
{\rm h}^{f}_{c,\Delta}\!
\left[^{\underline{\hspace{3pt}}\,\Delta_3
\;\underline{\hspace{3pt}}\,\Delta_2}_{\hspace{3pt}\,\Delta_4
\;\hspace{3pt}\, \Delta_1} \right]
+
\begin{array}[t]{c}
{\displaystyle\sum} \\[-6pt]
{\scriptscriptstyle
1 < rs \leq 2{f}}
\\[-8pt]
{\scriptscriptstyle
r + s\in 2{\mathbb N}
}
\end{array}
\frac{
{\mathcal R}^{{f}}_{c,\,rs}\!
\left[^{\underline{\hspace{3pt}}\,\Delta_3
\;\underline{\hspace{3pt}}\,\Delta_2}_{\hspace{3pt}\,\Delta_4
\;\hspace{3pt}\, \Delta_1} \right]
}
{
\Delta-\Delta_{rs}(c)
}\,,
\end{equation}
with $\Delta_{rs}(c)$ given by (\ref{delta:rs}),
or as a sum over the poles in $c:$
\begin{equation}
\label{first:expansion:c}
F^{f}_{c, \Delta}\!
\left[^{\underline{\hspace{3pt}}\,\Delta_3
\;\underline{\hspace{3pt}}\,\Delta_2}_{\hspace{3pt}\,\Delta_4
\;\hspace{3pt}\, \Delta_1} \right]
\; = \;
{\rm f}^{f}_{\Delta}\!
\left[^{\underline{\hspace{3pt}}\,\Delta_3
\;\underline{\hspace{3pt}}\,\Delta_2}_{\hspace{3pt}\,\Delta_4
\;\hspace{3pt}\, \Delta_1} \right]
+
\begin{array}[t]{c}
{\displaystyle\sum} \\[-6pt]
{\scriptscriptstyle
1< rs \leq 2{f},\ r > 1}
\\[-8pt]
{\scriptscriptstyle
r + s\in 2{\mathbb N}
}
\end{array}
\hspace*{-10pt}
\frac{
\widetilde{\mathcal R}^{{f}}_{\Delta,\,rs}\!
\left[^{\underline{\hspace{3pt}}\,\Delta_3
\;\underline{\hspace{3pt}}\,\Delta_2}_{\hspace{3pt}\,\Delta_4
\;\hspace{3pt}\, \Delta_1} \right]
}
{
c-c_{rs}(\Delta)
}\,,
\end{equation}
where $c_{rs}(\Delta)$ have the form (\ref{zero:rs}).

The function ${\rm h}^{f}_{c,\Delta}\!
\left[^{\underline{\hspace{3pt}}\,\Delta_3
\;\underline{\hspace{3pt}}\,\Delta_2}_{\hspace{3pt}\,\Delta_4
\;\hspace{3pt}\, \Delta_1} \right]$ can be determined
from the asymptotic behavior of $F^{f}_{c, \Delta}$ for large $\Delta,$
 while for ${\rm f}^{f}_{\Delta}\!
\left[^{\underline{\hspace{3pt}}\,\Delta_3
\;\underline{\hspace{3pt}}\,\Delta_2}_{\hspace{3pt}\,\Delta_4
\;\hspace{3pt}\, \Delta_1} \right]$ we simply have
\[
{\rm f}^{f}_{\Delta}\!
\left[^{\underline{\hspace{3pt}}\,\Delta_3
\;\underline{\hspace{3pt}}\,\Delta_2}_{\hspace{3pt}\,\Delta_4
\;\hspace{3pt}\, \Delta_1} \right]
\; = \;
\lim_{c\to\infty}
F^{f}_{c, \Delta}\!
\left[^{\underline{\hspace{3pt}}\,\Delta_3
\;\underline{\hspace{3pt}}\,\Delta_2}_{\hspace{3pt}\,\Delta_4
\;\hspace{3pt}\, \Delta_1} \right].
\]
Since $\Delta_{rs}\Big(c_{rs}(\Delta)\Big) = \Delta,$
the residue at $\Delta_{rs}(c)$ in (\ref{first:expansion:Delta}) and at $c_{rs}(\Delta)$ in (\ref{first:expansion:c})
are related by
\begin{eqnarray}
\nonumber
%\label{c:Delta:relation}
\widetilde{\mathcal R}^{{f}}_{\Delta,\,rs}\!
\left[^{\underline{\hspace{3pt}}\,\Delta_3
\;\underline{\hspace{3pt}}\,\Delta_2}_{\hspace{3pt}\,\Delta_4
\;\hspace{3pt}\, \Delta_1} \right]
&=&-
{\partial c_{rs}(\Delta)\over
\partial \Delta}\,
{\mathcal R}^{{f}}_{c_{rs}(\Delta),\,rs}\!
\left[^{\underline{\hspace{3pt}}\,\Delta_3
\;\underline{\hspace{3pt}}\,\Delta_2}_{\hspace{3pt}\,\Delta_4
\;\hspace{3pt}\, \Delta_1} \right],
\\[6pt]
\label{der}
{\partial c_{rs}(\Delta)\over \partial \Delta}
&=&
\frac{8c_{rs}(\Delta)-12}{\left(r^2-1\right)\beta_{rs}^4(\Delta) - \left(s^2-1\right)}.
\end{eqnarray}

We shall start with calculating the residue at $\Delta_{rs}$.
The corresponding pole
arises due to  the existence
of a singular vector
$
\chi_{rs} \in {\cal V}^{\frac{rs}{2}}_{\Delta_{rs}}.
$
The rank of the zero of the Kac determinant  at $\Delta
= \Delta_{rs}$ shows that in a generic case
({\em i.e.}\ when the supermodule ${\cal V}_{\Delta_{rs}+{rs\over 2}}$ is irreducible)
all the null vectors in ${\cal V}^{f}_{\Delta_{rs}}$ are descendants of $\chi_{rs}.$

Let $\chi_{rs}^{KM}$ be the coefficients of $\chi_{rs}$ in the basis $S_{-K}L_{-M}\nu_{\Delta_{rs}},$
\begin{equation}
\label{decomposition}
\chi_{rs} = \sum_{K,M}\chi_{rs}^{KM}S_{-K}L_{-M}\,\nu_{\Delta_{rs}}\,.
\end{equation}
We normalize $\chi_{rs}$ such that for $rs \in 2{\mathbb N}$ the coefficient at $(L_{-1})^{\frac{rs}{2}}\,\nu_{\Delta_{rs}}$,
and for $rs \in 2{\mathbb N}-1$ the coefficient at $S_{-\frac12}(L_{-1})^{\frac{rs-1}{2}}\,\nu_{\Delta_{rs}}$,
is equal 1.

\noindent
For ${f} > {rs\over 2}$ consider  vectors of the form
$$
S_{-K}L_{-M}\,\chi^\Delta_{rs}\in {\cal V}_\Delta^{f}\,,
\hskip 5mm
|K|+|M|= {f} - \textstyle {rs\over 2}\,,
$$
where
\[
\chi_{rs}^{\Delta} = \sum_{K,M}\chi_{rs}^{KM}S_{-K}L_{-M}\,\nu_{\Delta}\,,
\]
so that
$
\chi_{rs} = \lim_{\Delta\to \Delta_{rs}} \chi_{rs}^{\Delta}.
$
The set of these vectors can be always extended to a full basis in ${\cal V}_\Delta^{f}$.
Working in such a basis and using the properties of the Gram matrix $B^{f}_{c, \Delta}$
and its inverse one gets
\begin{eqnarray}
\label{res:1}
&&
{\mathcal R}^{{f}}_{c,\,rs}\!
\left[^{\underline{\hspace{3pt}}\,\Delta_3
\;\underline{\hspace{3pt}}\,\Delta_2}_{\hspace{3pt}\,\Delta_4
\;\hspace{3pt}\, \Delta_1} \right] \; = \; A_{rs}(c) \ \times\\
\nonumber
&&
\sum
    \rho^{\Delta_4\ \Delta_3 \ \Delta}_{\infty \ \ 1 \ \ \ 0} (\nu_4, \underline{\hspace{4pt}}\,\nu_3 , S_{-K}L_{-M}\chi_{rs} )
   \ \left[B^{{f}-\frac{rs}{2}}_{c, \Delta_{rs}+\frac{rs}{2}}\right]^{KM,LN} \!
   \rho^{\Delta\ \Delta_2 \ \Delta_1}_{\infty \ \ 1 \ \ \ 0}
   (S_{-L}L_{-N}\chi_{rs} ,  \underline{\hspace{4pt}}\,\nu_2 , \nu_1 ),
\end{eqnarray}
with
\begin{equation}
\label{A:rs:1}
A_{rs}(c)
\; = \;
\lim_{\Delta\to\Delta_{rs}}
\left(\frac{\left\langle\chi_{rs}^\Delta|\chi_{rs}^\Delta\right\rangle}{\Delta - \Delta_{rs}(c)}
\right)^{-1}.
\end{equation}
The factorization (\ref{factorization})
and  the reflection properties (\ref{reflection1}), (\ref{reflection2}) of the form $\rho$ give:
\begin{eqnarray}
\label{res:even}
{\mathcal R}^{{f}}_{c,\,rs}
\left[^{\underline{\hspace{3pt}}\,\Delta_3
\;\underline{\hspace{3pt}}\,\Delta_2}_{\hspace{3pt}\,\Delta_4
\;\hspace{3pt}\, \Delta_1} \right]
&=&
A_{rs}(c)\,S_{rs}(\underline{\hspace{4pt}}\,\Delta_3)\\
\nonumber
&&\hspace{-30pt}
\times\,\rho^{\Delta\ \Delta_3\ \Delta_4}_{\infty  \ 1 \ \ \ 0}\left(\chi_{rs},\underline{\hspace{4pt}}\,\nu_3,\nu_4\right)
\rho^{\Delta\ \Delta_2\ \Delta_1}_{\infty  \ 1 \ \ \ 0}\left(\chi_{rs},\underline{\hspace{4pt}}\,\nu_2,\nu_1\right)
F^{{f}-\frac{rs}{2}}_{c, \Delta_{rs} + \frac{rs}{2}}
\left[^{\underline{\hspace{3pt}}\,\Delta_3
\;\underline{\hspace{3pt}}\,\Delta_2}_{\hspace{3pt}\,\Delta_4
\;\hspace{3pt}\, \Delta_1} \right]
\end{eqnarray}
for ${f}-\frac{rs}{2} \in {\mathbb N}\cup \{0\}$,
and
\begin{eqnarray}
\label{res:odd}
{\mathcal R}^{{f}}_{c,\,rs}
\left[^{\underline{\hspace{3pt}}\,\Delta_3
\;\underline{\hspace{3pt}}\,\Delta_2}_{\hspace{3pt}\,\Delta_4
\;\hspace{3pt}\, \Delta_1} \right] &=&
A_{rs}(c)\,S_{rs}(\underline{\hspace{4pt}}\,\Delta_3)\\
\nonumber
&&\hspace{-30pt}
\rho^{\Delta\ \Delta_3\ \Delta_4}_{\infty  \ 1 \ \ \ 0}\left(\chi_{rs},\widetilde{\underline{\hspace{4pt}}\,\nu_3},\nu_4\right)
\rho^{\Delta\ \Delta_2\ \Delta_1}_{\infty  \ 1 \ \ \ 0}\left(\chi_{rs},\widetilde{\underline{\hspace{4pt}}\,\nu_2},\nu_1\right)
F^{{f}-\frac{rs}{2}}_{c, \Delta_{rs} + \frac{rs}{2}}
\left[^{\underline{\hspace{3pt}}\,\Delta_3
\;\underline{\hspace{3pt}}\,\Delta_2}_{\hspace{3pt}\,\Delta_4
\;\hspace{3pt}\, \Delta_1} \right]
\end{eqnarray}
for ${f}-\frac{rs}{2} \in {\mathbb N}-\frac12$,
where $\widetilde{\nu}= *\nu$, $\widetilde{*\nu} = \nu$, and
\begin{equation}
\label{S:rs}
S_{rs}(\Delta)=1\;\;\;,\;\;\;S_{rs}(*\Delta)=(-1)^{rs}.
\end{equation}

In order to calculate $\rho^{\Delta_3\ \Delta_2 \ \Delta_1}_{\infty \ \ 1 \ \ \ 0} (\chi_{rs}, \underline{\hspace{4pt}}\,\nu_2 , \nu_1 )$
we first observe that
by factorization (\ref{factorization}):
\begin{eqnarray}
\label{matrixele1'}
\varrho^{\Delta_3\ \Delta_2 \ \Delta_1}_{\infty \ \ 1 \ \ \ 0} (\chi_{rs}, \nu_2 , \nu_1 )
&=&
\rho^{\Delta_3\ \Delta_2 \ \Delta_1}_{\infty \ \ 1 \ \ \ 0} (\chi_{rs}, \nu_2 , \nu_1 )\times
\left\{
\begin{array}{lcl}
\varrho^{\Delta_{rs}\  \Delta_2 \ \Delta_1}_{\infty \ \ \ 1 \ \ \ 0}(\nu_{rs}, \nu_2 , \nu_1 ),
\\[12pt]
\varrho^{\Delta_{rs}\  \Delta_2 \ \Delta_1}_{\infty \ \ \ 1 \ \ \ 0} (\nu_{rs}, \ast \nu_2 , \nu_1 ),
%&& \frac{rs}{2} \in \mathbb{N}- \frac{1}{2}
\end{array}
\right.
\\[10pt]
\label{matrixele2'}
\varrho^{\Delta_{rs}\ \Delta_2 \ \Delta_1}_{\infty \ \ 1 \ \ \ 0}(\chi_{rs} , *\nu_2 , \nu_1 )
&=&
\rho^{\Delta_{rs}\ \Delta_2 \ \Delta_1}_{\infty \ \ 1 \ \ \ 0}(\chi_{rs} , *\nu_2 , \nu_1 )\times
\left\{
\begin{array}{lcl}
\varrho^{\Delta_{rs}\  \Delta_2 \ \Delta_1}_{\infty\ \ \ 1 \ \ \ 0}(\nu_{rs}, \ast \nu_2 , \nu_1 ),
\\[12pt]
\varrho^{\Delta_{rs}\  \Delta_2 \ \Delta_1}_{\infty\ \ \ 1 \ \ \ 0}(\nu_{rs}, \nu_2 , \nu_1 ),
\end{array}
\right.
\end{eqnarray}
where $rs \in 2{\mathbb N}$ in the upper lines, $rs \in 2{\mathbb N} + 1$ in the lower lines, and
$\nu_{rs} \equiv \nu_{\Delta_{rs}(c)}.$

The Feigin-Fuchs construction allows to represent
\(
\varrho^{\Delta_{rs}\  \Delta_2 \ \Delta_1}_{\infty \ \ \ 1 \ \ \ 0} (\nu_{rs}, \nu_2 , \nu_1 )
\)
as the left (chiral) part of the three point ``screened'' correlator
\cite{Bershadsky:1985dq}. It is non-zero provided  the weights
\(
\Delta_i  =  -\frac{1}{8}{\left(\beta - \frac{1}{\beta}\right)^2} + \frac{\alpha_i^2}{8}
\)
satisfy the {\em even fusion rule}:
\[
%\begin{equation}
%\label{fusion:even}
\alpha_2 \pm \alpha_1
\; = \;
(1-r+2k)\beta - (1-s+2l)\frac{1}{\beta}, \hskip 1cm k+l \in 2{\mathbb N} \cup \{0\}
%\end{equation}
\]
where $k,l$ are integers in the range \(0 \leq k \leq r-1,\ 0 \leq
l \leq s-1 \).
Similarly, one gets
\(
\varrho^{\Delta_{rs}\  \Delta_2 \ \Delta_1}_{\infty \ \ 1 \ \ \ 0} (\nu_{rs}, \ast \nu_2 , \nu_1 )\neq 0
\)
if and only if the {\em odd fusion rule}
\[
%\begin{equation}
%\label{fusion:odd}
\alpha_2 \pm \alpha_1
\; = \;
(1-r+2k)\beta - (1-s+2l)\frac{1}{\beta}, \hskip 1cm k+l \in 2{\mathbb N} -1,
%\end{equation}
\]
is satisfied. Since  the definition (\ref{unn:rho:def}) implies that for the null vector $\chi_{rs}$
\[
\varrho^{\Delta_3\ \Delta_2 \ \Delta_1}_{\infty \ \ 1 \ \ \ 0} (\chi_{rs},\underline{\hspace*{4pt}} \nu_2 , \nu_1 )
\; = \;
0,
\]
we conclude from (\ref{matrixele1'}) and (\ref{matrixele2'}) that also for the normalized form
\begin{equation}
\label{zeroes}
\rho^{\Delta_3\ \Delta_2 \ \Delta_1}_{\infty \ \ 1 \ \ \ 0} (\chi_{rs},\underline{\hspace*{4pt}} \nu_2 , \nu_1 )
\; = \;
0
\end{equation}
if the appropriate fusion rule is satisfied.

Let us define the ``fusion polynomials''
\begin{equation}
\label{P:1} P^{rs}_{c}\!\left[^{\Delta_2}_{\Delta_1}\right] \; =
\; \prod_{p=1-r}^{r-1} \prod_{q=1-s}^{s-1}
\left(\frac{\alpha_2-\alpha_1 +p\beta -
q\beta^{-1}}{2\sqrt2}\right) \left(\frac{\alpha_2+\alpha_1 +p\beta
- q\beta^{-1}}{2\sqrt2}\right)
\end{equation}
where $p+q -(r+s) \in 4{\mathbb Z} + 2$ ($p$ and $q$ are related
to the previously used variables through $p = r-1 -2k,\ q = s-1-2l$)
and
\begin{equation}
\label{P:2}
P^{rs}_{c}\!\left[^{*\Delta_2}_{\hspace{4pt}\Delta_1}\right] \; =
\; \prod_{p=1-r}^{r-1} \prod_{q=1-s}^{s-1}
\left(\frac{\alpha_2-\alpha_1 +p\beta -
q\beta^{-1}}{2\sqrt2}\right) \left(\frac{\alpha_2+\alpha_1 +p\beta
- q\beta^{-1}}{2\sqrt2}\right)
\end{equation}
with $p+q -(r+s) \in 4{\mathbb Z}.$ The following properties
can be easily obtained by simple combinatorics:
\begin{enumerate}
\item
$P^{rs}_{c}\!\left[^{\Delta_2}_{\Delta_1}\right]$ vanishes if the
even fusion rule is satisfied and
$P^{rs}_{c}\!\left[^{*\Delta_2}_{\hspace{4pt}\Delta_1}\right]$ if
odd fusion rule holds;
\item
$P^{rs}_{c}\!\left[^{\Delta_2}_{\Delta_1}\right]$ is a polynomial
of degree $\left[\frac{rs+1}{2}\right]$ in the variable $\Delta_2
- \Delta_1$ and of degree  $\left[\frac{rs+1}{4}\right]$ in
$\Delta_2 + \Delta_1$ and
$P^{rs}_{c}\!\left[^{*\Delta_2}_{\hspace{4pt}\Delta_1}\right]$ is
a polynomial of degree $\left[\frac{rs}{2}\right]$ in the
variable $\Delta_2 - \Delta_1$ and of degree
$\left[\frac{rs}{4}\right]$ in $\Delta_2 + \Delta_1;$
\item
coefficients of highest powers of $\Delta_2-\Delta_1$ in both
polynomials are equal 1.
\end{enumerate}
The properties above uniquely determine polynomials $P^{rs}_c.$
On the other hand, it follows from (\ref{zeroes}),
(\ref{matrixele1}), (\ref{matrixele2}) and  the normalization of
$\chi_{rs}$  that
with an appropriate choice of the second argument
the same properties are satisfied
 by
\(
\rho^{\Delta\ \Delta_3\ \Delta_4}_{\infty \ 1 \ \ \ 0}\left(\chi_{rs},\underline{\hskip 4pt}\nu_2,\nu_1\right)
\)
as well.
This implies the equalities:
\begin{eqnarray}
\label{rho:P} \nonumber
\rho^{\Delta\ \Delta_3\ \Delta_4}_{\infty \ 1 \ \ \ 0}\left(\chi_{rs},\nu_2,\nu_1\right)
& = &
\left\{
\begin{array}{lll}
P^{rs}_{c}\!\left[^{\Delta_2}_{\Delta_1}\right] &\;\;\;& {\rm
for}\ \frac{rs}{2} \in {\mathbb N},
\\[8pt]
P^{rs}_{c}\!\left[^{*\Delta_2}_{\hspace{4pt}\Delta_1}\right]
&\;\;\;& {\rm for}\ \frac{rs}{2} \in {\mathbb N}-\frac12,
\end{array}
\right.
\\
\\
\nonumber
\rho^{\Delta\ \Delta_3\ \Delta_4}_{\infty  \ 1 \ \ \ 0}\left(\chi_{rs},*\nu_2,\nu_1\right)
& = &
\left\{
\begin{array}{lll}
P^{rs}_{c}\!\left[^{*\Delta_2}_{\hspace{4pt}\Delta_1}\right]
&\;\;\;& {\rm for}\ \frac{rs}{2} \in {\mathbb N},
\\[8pt]
P^{rs}_{c}\!\left[^{\Delta_2}_{\Delta_1}\right] &\;\;\;& {\rm
for}\ \frac{rs}{2} \in {\mathbb N}-\frac12.
\end{array}
\right.
\end{eqnarray}

In order to complete our derivation of
\( {\mathcal
R}^{{f}}_{c,\,rs} \left[^{\underline{\hspace{3pt}}\,\Delta_3
\;\underline{\hspace{3pt}}\,\Delta_2}_{\hspace{3pt}\,\Delta_4
\;\hspace{3pt}\, \Delta_1} \right]
\)
we thus only need the
coefficients $A_{rs}(c).$  Fortunately the r.h.s.\ of
(\ref{A:rs:1}) can be easily calculated using equations (43) -- (46) of
\cite{Belavin:2006} (note that in the present case we calculate
the residue at $\Delta_{rs}$ rather than at $\alpha_{rs}$). In our notation the
result reads:
\begin{equation}
\label{A:rs:2}
A_{rs}(c)
\; = \;
\frac12 (-1)^{rs-1}
\prod_{p=1-r}^r
\prod_{q=1-s}^s
\left(\frac{1}{\sqrt{2}}\left(p\beta -\frac{q}{\beta}\right)\right)^{-1}
\hskip -3mm,
\hskip .5cm
p+q \in 2{\mathbb Z}, \; (p,q) \neq (0,0),(r,s).
\end{equation}
\vskip 5mm

\noindent
Our final formulae for the residue at $\Delta_{rs}$ take the form:
\begin{eqnarray}
\label{res:evenf}
{\mathcal R}^{m}_{c,\,rs}\!
\left[^{\underline{\hspace{3pt}}\,\Delta_3
\;\underline{\hspace{3pt}}\,\Delta_2}_{\hspace{3pt}\,\Delta_4
\;\hspace{3pt}\, \Delta_1} \right]
&=&
A_{rs}(c)\,S_{rs}(\underline{\hspace{4pt}}\,\Delta_3)\,
P^{rs}_{c}\!\left[^{\underline{\hspace{3pt}}\,\Delta_3}_{\hspace{5pt}\Delta_4}\right]
P^{rs}_{c}\!\left[^{\underline{\hspace{3pt}}\,\Delta_2}_{\hspace{5pt}\Delta_1}\right]
F^{m-\frac{rs}{2}}_{c, \Delta_{rs} + \frac{rs}{2}}\!
\left[^{\underline{\hspace{3pt}}\,\Delta_3
\;\underline{\hspace{3pt}}\,\Delta_2}_{\hspace{3pt}\,\Delta_4
\;\hspace{3pt}\, \Delta_1} \right]
\end{eqnarray}
for $m \in {\mathbb N}\cup \{0\}$ and
\begin{eqnarray}
\label{res:oddf}
{\mathcal R}^{k}_{c,\,rs}\!
\left[^{\underline{\hspace{3pt}}\,\Delta_3
\;\underline{\hspace{3pt}}\,\Delta_2}_{\hspace{3pt}\,\Delta_4
\;\hspace{3pt}\, \Delta_1} \right] &=&
A_{rs}(c)\,S_{rs}(\underline{\hspace{4pt}}\,\Delta_3)\,
P^{rs}_{c}\!\left[^{\widetilde{\underline{\hspace{3pt}}\,\Delta_3}}_{\hspace{5pt}\Delta_4}\right]
P^{rs}_{c}\!\left[^{\widetilde{\underline{\hspace{3pt}}\,\Delta_2}}_{\hspace{5pt}\Delta_1}\right]
F^{k-\frac{rs}{2}}_{c, \Delta_{rs} + \frac{rs}{2}}\!
\left[^{\underline{\hspace{3pt}}\,\Delta_3
\;\underline{\hspace{3pt}}\,\Delta_2}_{\hspace{3pt}\,\Delta_4
\;\hspace{3pt}\, \Delta_1} \right]
\end{eqnarray}
for $k \in {\mathbb N}- \frac12,$ where
the coefficients involved are defined by (\ref{S:rs}),(\ref{P:1}),(\ref{P:2}), and (\ref{A:rs:2}).

Let us now calculate the leading terms in the expansion over the poles in $c,$
\begin{equation}
\label{asymptotics}
{\rm f}^{f}_{\Delta}\!
\left[^{\underline{\hspace{3pt}}\,\Delta_3\;\underline{\hspace{3pt}}\,\Delta_2}_{\hspace{3pt}\,\Delta_4\;\hspace{3pt}\, \Delta_1} \right]
\; = \;
\lim_{c \to\infty}
F^{f}_{c, \Delta}\!
\left[^{\underline{\hspace{3pt}}\,\Delta_3\;\underline{\hspace{3pt}}\,\Delta_2}_{\hspace{3pt}\,\Delta_4\;\hspace{3pt}\, \Delta_1} \right].
\end{equation}
It follows from the $NS$ algebra  that the only vectors in ${\cal V}^n_{c,\Delta}$
with all matrix elements independent of $c$ (an thus finite in the limit $c\to\infty$)
are proportional to $L_{-1}^n\ket{\Delta}$. For ${\cal V}^{n+\frac12}_{c,\Delta}$
such vectors are proportional to  $S_{-\frac12}L_{-1}^n\ket{\Delta}.$
Consequently, only these states contribute to (\ref{asymptotics}). We have:
\begin{eqnarray}
\label{matrix:elements}
\nonumber
\bra{\Delta}L_1^n L_{-1}^n \ket{\Delta}
& = &
n!\big(2\Delta\big)_n,
\\
\bra{\Delta}L_1^n S_{\frac12}  S_{-\frac12}L_{-1}^n \ket{\Delta}
& = &
n!\big(2\Delta\big)_{n+1},
\end{eqnarray}
where
\[
\big(a\big)_n = \frac{\Gamma(a+n)}{\Gamma(n)}
\]
is the Pochhammer symbol.

In the limit $c\to\infty$ the corresponding elements of the inverse Gram matrix
are just inverses of the matrix elements (\ref{matrix:elements}). Furthermore,
from (\ref{matrixele1}) and (\ref{matrixele2}):
\begin{eqnarray*}
\rho^{\Delta\ \Delta_2 \ \Delta_1}_{\infty \ \ 1 \ \ \ 0} \left(\mathrm{L}_{-1}^n \nu, \nu_2 , \nu_1 \right)
& = &
\big(\Delta+\Delta_2-\Delta_1\big)_n,
\\
\rho^{\Delta\ \Delta_2 \ \Delta_1}_{\infty \ \ 1 \ \ \ 0} \left(S_{-\frac12}\mathrm{L}_{-1}^n \nu, \nu_2 , \nu_1 \right)
& = &
\left(\Delta+\Delta_2-\Delta_1+\frac12\right)_n,
\\
\rho^{\Delta\ \Delta_2 \ \Delta_1}_{\infty \ \ 1 \ \ \ 0} \left(\mathrm{L}_{-1}^n \nu, *\nu_2 , \nu_1 \right)
& = &
\left(\Delta+\Delta_2-\Delta_1+\frac12\right)_n,
\\
\rho^{\Delta\ \Delta_2 \ \Delta_1}_{\infty \ \ 1 \ \ \ 0} \left(S_{-\frac12}\mathrm{L}_{-1}^n \nu, *\nu_2 , \nu_1 \right)
& = &
-\rho^{\Delta_1\ \Delta_2 \ \Delta}_{0 \ \ \ 1 \ \ \infty} \left(\nu_ 1, *\nu_2 , S_{-\frac12}\mathrm{L}_{-1}^n \nu \right)
\; = \;
\big(\Delta+\Delta_2-\Delta_1\big)_{n+1}.
\end{eqnarray*}
We thus  get:
\begin{eqnarray}
\label{asymptotics:final}
\nonumber
{\rm f}^{n}_{\Delta}\! \left[^{ \Delta_3 \ \Delta_2}_{ \Delta_4 \  \Delta_1}\right]
& = &
\frac{1}{n!}
\frac{
\big(\Delta+\Delta_3-\Delta_4\big)_{n}\big(\Delta+\Delta_2-\Delta_1\big)_{n}
}{
\big(2\Delta\big)_n
},
\\
\nonumber
{\rm f}^{n + \frac12}_{\Delta}\! \left[^{ \Delta_3 \ \Delta_2}_{ \Delta_4 \  \Delta_1}\right]
& = &
\frac{1}{n!}
\frac{
\left(\Delta+\Delta_3-\Delta_4+\frac12\right)_{n}\left(\Delta+\Delta_2-\Delta_1+\frac12\right)_{n}
}{
\big(2\Delta\big)_{n+1}
},
\\
{\rm f}^{n }_{\Delta}\! \left[^{*\Delta_3 \ *\Delta_2}_{\ \Delta_4 \ \ \Delta_1}\right]
& = &
\frac{1}{n!}
\frac{
\left(\Delta+\Delta_3-\Delta_4+\frac12\right)_{n}\left(\Delta+\Delta_2-\Delta_1+\frac12\right)_{n}
}{
\big(2\Delta\big)_{n}
},
\\
\nonumber
{\rm f}^{n + \frac12}_{\Delta}\! \left[^{*\Delta_3 \ *\Delta_2}_{\ \Delta_4 \ \ \Delta_1}\right]
& = &
- \
\frac{1}{n!}
\frac{
\left(\Delta+\Delta_3-\Delta_4\right)_{n+1}\big(\Delta+\Delta_2-\Delta_1\big)_{n+1}
}{
\big(2\Delta\big)_{n+1}
},
\end{eqnarray}
and so on.

Substituting (\ref{der}), (\ref{res:evenf}), (\ref{res:oddf}) and  (\ref{asymptotics:final}) into
(\ref{first:expansion:c}),
and introducing simplified notation
\begin{equation}
\label{tildas}
\tilde A_{rs}(\Delta) \; = \; -{\partial c_{rs}(\Delta)\over
\partial \Delta}\,A_{rs}\Big(c_{rs}(\Delta)\Big), \hskip 5mm
P^{rs}_{\Delta}\!\left[^{\underline{\hspace{3pt}}\,\Delta_a}_{\hspace{5pt}\Delta_b}\right]
\; = \;
P^{rs}_{c_{rs}(\Delta)}\!\left[^{\underline{\hspace{3pt}}\,\Delta_a}_{\hspace{5pt}\Delta_b}\right],
\end{equation}
one finally gets the recursion relations for the coefficients in the $x$-expansion of the NS superconformal blocks
\begin{eqnarray}
\label{recursion:even}
F^m_{c, \Delta}\!
\left[^{\underline{\hspace{3pt}}\,\Delta_3
\;\underline{\hspace{3pt}}\,\Delta_2}_{\hspace{3pt}\,\Delta_4
\;\hspace{3pt}\, \Delta_1} \right]
&\ =\ &
{\rm f}^m_{\Delta}\!
\left[^{\underline{\hspace{3pt}}\,\Delta_3
\;\underline{\hspace{3pt}}\,\Delta_2}_{\hspace{3pt}\,\Delta_4
\;\hspace{3pt}\, \Delta_1} \right]
\\
\nonumber
& + &
\hspace*{-20pt}
\begin{array}[t]{c}
{\displaystyle\sum} \\[-6pt]
{\scriptscriptstyle
1< rs \leq 2{f},\ 1<r }
\\[-8pt]
{\scriptscriptstyle
r + s\in 2{\mathbb N}}
\end{array}
%ssssssssssssssssssssss
{\tilde A_{rs}(\Delta)
%S_{rs}(\underline{\hspace{4pt}}\,\Delta_3)
\over
c- c_{rs}(\Delta)
}\
 P^{rs}_{\Delta}\!\left[^{\underline{\hspace{3pt}}\,\Delta_3}_{\hspace{5pt}\Delta_4}\right]
P^{rs}_{\Delta}\!\left[^{\underline{\hspace{3pt}}\,\Delta_2}_{\hspace{5pt}\Delta_1}\right]
F^{m-\frac{rs}{2}}_{c_{rs}, \Delta + \frac{rs}{2}}
\left[^{\underline{\hspace{3pt}}\,\Delta_3
\;\underline{\hspace{3pt}}\,\Delta_2}_{\hspace{3pt}\,\Delta_4
\;\hspace{3pt}\, \Delta_1} \right],
\end{eqnarray}
where $m\in \mathbb{N}$ and
\begin{eqnarray}
\label{recursion:even:2}
F^k_{c, \Delta}\!
\left[^{\underline{\hspace{3pt}}\,\Delta_3
\;\underline{\hspace{3pt}}\,\Delta_2}_{\hspace{3pt}\,\Delta_4
\;\hspace{3pt}\, \Delta_1} \right]
& = &
{\rm f}^k_{\Delta}\!
\left[^{\underline{\hspace{3pt}}\,\Delta_3
\;\underline{\hspace{3pt}}\,\Delta_2}_{\hspace{3pt}\,\Delta_4
\;\hspace{3pt}\, \Delta_1} \right]
\\
\nonumber
&\pm &
\hspace*{-15pt}
\begin{array}[t]{c}
{\displaystyle\sum} \\[-6pt]
{\scriptscriptstyle
1< rs \leq 2{f},\ 1<r}
\\[-8pt]
{\scriptscriptstyle
r + s\in 2{\mathbb N}}
\end{array}
%ssssssssssssssssssssssss
{\tilde A_{rs}(\Delta)
%S_{rs}(\underline{\hspace{4pt}}\,\Delta_3)
\over
c- c_{rs}(\Delta)
}\
 P^{rs}_{\Delta}\!\left[^{\widetilde{\underline{\hspace{3pt}}\,\Delta_3}}_{\hspace{5pt}\Delta_4}\right]
 P^{rs}_{\Delta}\!\left[^{\widetilde{\underline{\hspace{3pt}}\,\Delta_2}}_{\hspace{5pt}\Delta_1}\right]
F^{k-\frac{rs}{2}}_{c_{rs}, \Delta + \frac{rs}{2}}
\left[^{\underline{\hspace{3pt}}\,\Delta_3
\;\underline{\hspace{3pt}}\,\Delta_2}_{\hspace{3pt}\,\Delta_4
\;\hspace{3pt}\, \Delta_1} \right],
\end{eqnarray}
where $k\in \mathbb{N}-{1\over 2}$, and $+$/$-$ in the second line correspond to $\Delta_3$/$*\Delta_3$, respectively.
Let us note that for each type of the NS superconformal block one gets independent recursion formulae mixing coefficients of
the even
and the odd blocks.

\section*{Acknowledgements}
The work of L.H and Z.J. was partially supported by the Polish State Research
Committee (KBN) grant no.\ 1 P03B 025 28.

\vskip 1mm
\noindent
The research of L.H.\ is supported by the Alexander von Humboldt Foundation scholarship.

\vskip 1mm
\noindent
L.H.\ would like to thank Alexander Belavin for sharing with him his interest in the topics discussed
in the present work and numerous helpful discussions at the early stage of the project.\\
P.S.\ is grateful to the faculty of the Institute of Theoretical Physics, University of Wroc\l{}aw,
for the hospitality.

\section*{Appendix A. Explicit calculations at $f=\frac32, 2,$ and $\frac52$}

In this appendix we present some explicit calculation for the first few coefficients
of ``symmetric'' conformal blocks. Our purpose is to illustrate how the presented method works
and to provide a simple check of the  formulae derived.

At $f = \frac12$ and $f = 1$ there are no  $c$ poles in the Gram matrix and the coefficients
of the blocks are fully determined by the asymptotic terms (\ref{asymptotics:final}).

For $f = \frac32$ there is a pole at
\[
c = c_{31}(\Delta) = \frac{3\Delta(3-2\Delta)}{2\Delta+1}
\hskip 5mm
\Rightarrow
\hskip 5mm
-\frac{\partial c_{31}(\Delta)}{\partial\Delta}
=  3\frac{(2\Delta-1)(2\Delta+3)}{(2\Delta+1)^2}.
\]
We have:
\begin{eqnarray*}
 F^{\frac32}_{c, \Delta}\! \left[^{*\Delta_3 \ *\Delta_2}_{ \ \Delta_4 \ \ \Delta_1} \right]
&=&
 {\rm f}^{\frac32}_{\Delta}\! \left[^{*\Delta_3 \ *\Delta_2}_{ \ \Delta_4 \ \ \Delta_1} \right]
-
\frac{\tilde A_{31}(\Delta)}{c-c_{31}}\
P^{31}_\Delta\left[^{\Delta_2}_{\Delta_1}\right]
P^{31}_\Delta\left[^{\Delta_3}_{\Delta_4}\right],
\\
\\
F^{\frac32}_{c, \Delta}\! \left[^{ \Delta_3 \ \Delta_2}_{ \Delta_4 \  \Delta_1}\right]
&=&
{\rm f}^{\frac32}_{\Delta}\! \left[^{ \Delta_3 \ \Delta_2}_{ \Delta_4 \  \Delta_1}\right]
+\
\frac{\tilde A_{31}(\Delta)}{c-c_{31}}\
P^{31}_\Delta\left[^{*\Delta_2}_{\ \Delta_1}\right]
P^{31}_\Delta\left[^{*\Delta_3}_{\ \Delta_4}\right],
\end{eqnarray*}
with (from (\ref{P:1}), (\ref{P:2}) and (\ref{tildas}))
\begin{eqnarray*}
P^{31}_{\Delta}\left[^{\Delta_2}_{\Delta_1}\right]
& = &
\frac12
\Big(
2(\Delta_2-\Delta_1)^2 - (2\Delta+1)(\Delta_2 + \Delta_1) + \Delta
\Big),
\\
P^{31}_{\Delta}\left[^{*\Delta_2}_{\ \Delta_1}\right]
& = &
(\Delta_2-\Delta_1),
\end{eqnarray*}
and (from (\ref{A:rs:2}) and (\ref{tildas}))
\[
\tilde A_{31}(\Delta) \; = \; \frac{6}{(2\Delta+1)^2}.
\]
These formulae are easily checked using the definitions from the section
\ref{definitions} and an explicit expression for the Gram matrix, which
in the basis $\left\{S_{-\frac12}L_{-1}\ket{\Delta},\ S_{-\frac32}\ket{\Delta}\right\}$ reads:
\[
\left[B^{{\frac32}}_{c,\Delta}\right]
\; = \;
\left(
\begin{array}{cc}
2\Delta(2\Delta +1) & 4\Delta
\\
4\Delta & 2\Delta + \frac{2}{3}c
\end{array}
\right).
\]

For $f=2$ we have poles at $c = c_{31}(\Delta)$ and
\[
c = c_{22}(\Delta) = \frac{3}{2} - 8\Delta.
\]
With
\begin{eqnarray*}
P^{22}_{\Delta}\left[^{\Delta_2}_{\Delta_1}\right]
& = &
\frac13
\Big(
3(\Delta_2-\Delta_1)^2 -2\Delta(\Delta_2 + \Delta_1) - \Delta^2
\Big),
\\
P^{22}_{\Delta}\left[^{*\Delta_2}_{\ \Delta_1}\right]
& = &
\frac{1}{12}
\Big(
12(\Delta_2 -\Delta_1)^2 -4(2\Delta+3)(\Delta_2 + \Delta_1) - (2\Delta+3)(2\Delta-1)
\Big),
\end{eqnarray*}
and
\[
\tilde A_{22}(\Delta) \; = \; \frac{9}{4\Delta(2\Delta +3)}\,,
\]
we have:
\begin{eqnarray*}
 F^{2}_{c, \Delta}\! \left[^{ \Delta_3 \  \Delta_2}_{ \Delta_4 \ \Delta_1}\right]
\; = \;
{\rm f}^{2}_{\Delta}\! \left[^{ \Delta_3 \  \Delta_2}_{ \Delta_4 \ \Delta_1}\right]
& + &  \
\frac{\tilde A_{22}(\Delta)}{c-c_{22}}\
P^{22}_\Delta\left[^{\Delta_2}_{\Delta_1}\right]
P^{22}_\Delta\left[^{\Delta_3}_{\Delta_4}\right]
\\
& + & \
    \frac{\tilde A_{31}(\Delta)}{c-c_{31}}\
    P^{31}_\Delta\left[^{\Delta_2}_{\Delta_1}\right]
    P^{31}_\Delta\left[^{\Delta_3}_{\Delta_4}\right]
    {\rm f}^{\frac12}_{\Delta+ \frac32}\! \left[^{ \Delta_3 \  \Delta_2}_{ \Delta_4 \ \Delta_1}\right],
\end{eqnarray*}
\begin{eqnarray*}
F^{2}_{c, \Delta}\! \left[^{*\Delta_3 \ * \Delta_2}_{\ \Delta_4 \ \ \Delta_1}\right]
\; = \;
{\rm f}^{2}_{\Delta}\! \left[^{ *\Delta_3 \ * \Delta_2}_{\ \Delta_4 \ \ \Delta_1}\right]
&+& \
\frac{\tilde A_{22}(\Delta)}{c-c_{22}}\
P^{22}_\Delta\left[^{*\Delta_2}_{\ \Delta_1}\right]
P^{22}_\Delta\left[^{*\Delta_3}_{\ \Delta_4}\right]
\\
&+& \
    \frac{\tilde A_{31}(\Delta)}{c-c_{31}}\
    P^{31}_\Delta\left[^{*\Delta_2}_{\ \Delta_1}\right]
    P^{31}_\Delta\left[^{*\Delta_3}_{\ \Delta_4}\right]
    {\rm f}^{\frac12}_{\Delta+ \frac32}\! \left[^{* \Delta_3 \ *\Delta_2}_{ \ \Delta_4  \ \  \Delta_1}\right],
\end{eqnarray*}
what can be checked noting that in the basis
$\left\{
L_{-1}L_{-1}\ket{\Delta},\
L_{-2}\ket{\Delta},\
S_{-\frac12}S_{-\frac32}\ket{\Delta}
\right\}$
the Gram matrix reads
\[
\left[B^2_{c,\Delta}\right]
\; = \;
\left(
\begin{array}{cccc}
4\Delta(2\Delta +1) & 6\Delta && 8\Delta
\\
6\Delta & 4\Delta + \frac{c}{2} && 3\Delta+c
\\
8\Delta & 3\Delta+c & \ &4\Delta\left(\Delta+\frac{c}{3}\right) +2 (c-\Delta)
\end{array}
\right).
\]

Finally, for $f=\frac{5}{2}$ there appears a ``new'' pole at
\[
c = c_{51}(\Delta) = \frac{(2\Delta -1)(5-\Delta)}{2\Delta + 2}
\hskip 5mm
\Rightarrow
\hskip 5mm
-\frac{\partial c_{51}(\Delta)}{\partial\Delta}
\; = \;
\frac{(\Delta-2)(\Delta+4)}{(\Delta+1)^2}.
\]
The fusion polynomials read
\begin{eqnarray*}
P^{51}_{\Delta}\left[^{\Delta_2}_{\Delta_1}\right]
& = &
\frac13(\Delta_2-\Delta_1)
\Big(
3(\Delta_2-\Delta_1)^2 -4(\Delta+1)(\Delta_2+\Delta_1) + \Delta(\Delta+4)
\Big),
\\
P^{51}_{\Delta}\left[^{*\Delta_2}_{\ \Delta_1}\right]
& = &
\frac{1}{12}
\Big(
12(\Delta_2-\Delta_1)^2 -4(\Delta+1)(\Delta_2+\Delta_1) + 2\Delta-1
\Big),
\end{eqnarray*}
and with
\[
\tilde A_{51}(\Delta) = \frac{9}{8\Delta(\Delta +2)(\Delta+1)^2}
\]
we have:
\begin{eqnarray*}
 F^{\frac52}_{c, \Delta}\! \left[^{*\Delta_3 \ * \Delta_2}_{\  \Delta_4 \ \ \Delta_1}\right]
\; = \;
{\rm f}^{\frac52}_{\Delta}\! \left[^{*\Delta_3 \ *\Delta_2}_{\  \Delta_4 \ \ \Delta_1}\right]
& - &\
\frac{\tilde A_{51}(\Delta)}{c-c_{51}}\
P^{51}_\Delta\left[^{\Delta_2}_{\Delta_1}\right]
P^{51}_\Delta\left[^{\Delta_3}_{\Delta_4}\right]
\\
& + &\
\frac{\tilde A_{22}(\Delta)}{c-c_{22}}\
P^{22}_\Delta\left[^{\Delta_2}_{\Delta_1}\right]
P^{22}_\Delta\left[^{\Delta_3}_{\Delta_4}\right]
    {\rm f}^{\frac12}_{\Delta+ 2}\! \left[^{ *\Delta_3 \ * \Delta_2}_{\ \Delta_4 \ \ \Delta_1}\right]
\\
& - &\
    \frac{\tilde A_{31}(\Delta)}{c-c_{31}}\
    P^{31}_\Delta\left[^{\Delta_2}_{\Delta_1}\right]
    P^{31}_\Delta\left[^{\Delta_3}_{\Delta_4}\right]
    {\rm f}^{1}_{\Delta+ \frac32}\!\left[^{ *\Delta_3 \ *\Delta_2}_{ \  \Delta_4 \ \  \Delta_1}\right],
\end{eqnarray*}
\begin{eqnarray*}
F^{\frac52}_{c, \Delta}\! \left[^{ \Delta_3 \ \Delta_2}_{ \Delta_4 \  \Delta_1}\right]
\; = \;
{\rm f}^{\frac52}_{\Delta}\! \left[^{ \Delta_3 \ \Delta_2}_{ \Delta_4 \  \Delta_1}\right]
& + & \
\frac{\tilde A_{51}(\Delta)}{c-c_{51}}\
P^{51}_\Delta\left[^{*\Delta_2}_{\ \Delta_1}\right]
P^{51}_\Delta\left[^{*\Delta_3}_{\ \Delta_4}\right]
\\
& + &\
\frac{\tilde A_{22}(\Delta)}{c-c_{22}}\
P^{22}_\Delta\left[^{*\Delta_2}_{\ \Delta_1}\right]
P^{22}_\Delta\left[^{*\Delta_3}_{\ \Delta_4}\right]
{\rm f}^{\frac12}_{\Delta+2}\! \left[^{ \Delta_3 \ \Delta_2}_{ \Delta_4 \  \Delta_1}\right]
\\
&+&\
    \frac{\tilde A_{31}(\Delta)}{c-c_{31}}\
    P^{31}_\Delta\left[^{*\Delta_2}_{\ \Delta_1}\right]
    P^{31}_\Delta\left[^{*\Delta_3}_{\ \Delta_4}\right]
    {\rm f}^{1}_{\Delta+ \frac32}\! \left[^{ \Delta_3 \  \Delta_2}_{ \Delta_4 \  \Delta_1}\right].
\end{eqnarray*}
The results above fully agree with the direct calculation in which the Gram matrix,
with the form
\[
\left[B^{\frac52}_{c,\Delta}\right]
=
\left(
\begin{array}{cccc}
8\Delta(\Delta+1)(2\Delta +1) &  12\Delta(\Delta+1)  & 4\Delta(4\Delta+1)  & 12\Delta
\\
12\Delta(\Delta+1) & (\Delta+2)(8\Delta+c) - \frac32(3\Delta+c) & 7\Delta & 8\Delta + c
\\
4\Delta(4\Delta+1) & 7\Delta & \frac43 \Delta(3\Delta +3+c) & 4\Delta
\\
12\Delta & 8\Delta + c & 4\Delta & 2\Delta + 2c
\end{array}
\right)
\]
in the basis \(
\left\{
S_{-\frac12}L_{-1}L_{-1}\ket{\Delta},
\
S_{-\frac12}L_{-2}\ket{\Delta},
\
S_{-\frac32}L_{-1}\ket{\Delta},
\
S_{-\frac52}\ket{\Delta}
\right\},
\)
is used.

\end{document}